\documentstyle [12pt]{article}

\newcommand{\ds}{{\prime\prime}}
\newcommand{\tp}{{\prime\prime\prime}}
\newcommand{\ol}{\overline}
\newcommand{\ul}{\underline}
\newcommand{\be}{\begin{equation}}
\newcommand{\ee}{\end{equation}}
\newcommand{\ba}{\begin{array}}
\newcommand{\ea}{\end{array}}
\newcommand{\tls}{\theta^\star_L}
\newcommand{\trs}{\theta^\star_R}
\newcommand{\th}{\theta}
\newcommand{\tr}{{\rm tr}}
\newcommand{\Tr}{{\rm Tr}}
\newcommand{\vier}{\\ [4 pt]}
\newcommand{\acht}{\\ [12 pt]}
\newcommand{\tha}{\theta^\ast}
\newcommand{\tht}{\tilde\theta}

\newcommand{\cero}{\;\;\; 0}
\newcommand{\un}{\underline}

\begin{document}
\begin{titlepage}
\title{
\hfill {\normalsize SCIPP 96/10} \\ 
\hfill \\ \hfill \\
Asymmetric Orbifolds and Higher Level Models}

\vspace{5.cm}

\author{ \\
{\sc Jens Erler}\thanks{e-mail: erler@scipp.ucsc.edu} \\ \hfill \\ 
{\small Santa Cruz Institute for Particle Physics} \\
{\small University of California, Santa Cruz, CA 95064 (USA)} \\ } 

\maketitle
\thispagestyle{empty}

\begin{abstract}
\noindent
I introduce a class of string constructions based on asymmetric orbifolds 
leading to level two models. In particular, I derive in detail various
models with gauge groups $E_6$ and $SO(10)$, including a four generation
$E_6$ model with two adjoint representations. The occurrence of multiple
adjoint representations is a generic feature of the construction.
In the course of describing this approach, I will address the problem of 
twist phases in higher twisted sectors of asymmetric orbifolds.
\end{abstract}
\thispagestyle{empty}
\end{titlepage}

\setcounter{page}{1} 

\section{Introduction}
The apparent unification of gauge couplings~\cite{LP} in the context of
minimal supersymmetry when extrapolated to high energies has
created growing interest in supersymmetric Grand Unified Theories 
and Superstring Theories. The unification scale is impressively
close to but slightly lower than the string scale~\cite{DKL}. 
From a string theorist point of view, it is presently unclear whether 
this is indicative for an intermediate Grand Unification (GUT) group or
rather for string effects such as threshold~\cite{DKL,MNS} or strong
coupling effects~\cite{Witten}. 

If one chooses to invoke a GUT group such as $SO(10)$, it is necessary to 
construct string models based on level $k>1$ Kac-Moody algebras in order 
to potentially obtain Higgs representations able to break the gauge group 
in a satisfactory way. This, however, is rather awkward and most of the
string models constructed so far are typically realized at level 1. 
Moreover, all known leptons and quarks are either singlets or in the 
fundamental representations of $SU(3)$ and $SU(2)$ which are already
available at level 1. Similarly, non-standard Higgs representations
such as triplets are strongly constrained as they yield tree-level
contributions to the $\rho_0$-parameter\footnote{By definition, 
$\rho_0$ describes new physics contributions to the $\rho$ parameter so that
in the Standard Model with its minimal Higgs sector $\rho_0 \equiv 1$.}.
Indeed, through high precision measurements, $\rho_0$ is now known to be 
very close to unity\footnote{This value was obtained allowing even extra 
parameters describing non-standard loop contributions to the vector boson
self-energies ($S$, $T$ and $U$) and non-standard couplings of the $Z$ boson 
to $b$ quarks.}~\cite{EL},
\be
              \rho_0 = 0.9985 \pm 0.0019^{+0.0012}_{-0.0009}.
\ee
Although all this seems to encourage the construction of level 1 models
and consequently the rejection of a simple intermediate GUT group, in such
a case one necessarily encounters phenomenologically problematic fractionally 
charged particles\footnote{All Standard Model particles have integral charges
in the sense that if they transform in the singlet (triplet, antitriplet) 
congruency class of $SU(3)_C$ then they have electric charge 0 (2/3, 1/3) mod 
{\bf Z}.}. This important statement has been made precise by 
Schellekens~\cite{Schellekens}: any level 1 compactification of the
heterotic string with the GUT scale normalization of $\sin^2 \theta_W = 3/8$
has either fractional charges\footnote{As has been noted in 
Reference~\cite{CCHL}, there remains the logical possibility that the 
fractionally charged states appear only at massive levels, but no examples 
are known.} or an enhanced gauge group containing $SU(5)$.

The remaining options are either to accept fractional charges and to make them
sufficiently heavy and rare~\cite{AADF} or to confine them through an extra
non-Abelian gauge group~\cite{AEHN}, or alternatively, to proceed to higher 
levels $k$. The latter is the option chosen in this work. Since higher level
models generically (though not always) possess adjoint representations, it is
natural to use them for GUT breaking. However, this is not the only 
possibility and one may use adjoints in a rather different way: it has been 
noted that the addition of a color-octet (iso-singlet) and an iso-triplet 
(color-singlet) to the minimal supersymmetric standard model (MSSM)
can lead to gauge coupling unification at the correct (string) scale
with a lower $\alpha_s$ (as found in most low-energy determinations), when
the masses of both extra multiplets are a few times $10^{12}$ GeV~\cite{BFY}.

Higher level models were discussed
systematically by Lewellen~\cite{Lewellen} who also presented
some level 2 examples using free fermions. The fermionic construction
was then further exploited in References~\cite{CCHL,Cleaver}. Level 1
models with gauge groups $G\times G$, which can be broken to the diagonal
$G$ by turning on flat directions were studied by Finnell~\cite{Finnell},
who found three generation $SU(5)$ models of this type. The diagonal $G$ is 
then expected~\cite{FIQ} to be realized at level 2.

Various methods in the context of symmetric orbifolds were introduced in 
Reference~\cite{FIQ} and recently followed up~\cite{AFIU}. In the present
paper, I focus on asymmetric $Z_2 \times Z_N$ orbifolds which lead to
level 2 Kac-Moody algebras. This includes, in particular, models with
gauge group $E_6$. This group has the unique property of being able
to accommodate each fermion generation in its chiral fundamental 
representation. $E_6$ string GUT models have not been constructed before. They
have the special feature that no unwanted exotic representations
can occur in the massless spectrum. In the construction here,
the exceptional groups appear quite naturally: the simplest version of
this construction (see section~\ref{asymmetric}) yields $E_8$ and
$N=2$ supersymmetry (or simple supersymmetry in $D=6$); the simplest
$N=1$ model has gauge group $E_7$; and among the simplest {\em chiral\/} 
possibilities is $E_6$. 

This article is organized as follows: Section~\ref{general} summarizes some 
facts about higher level string models. In section~\ref{asymmetric}, I discuss 
the relevant aspects of asymmetric orbifolds and introduce the basic 
strategy how to obtain models at level 2. Section~\ref{E6} describes 
$Z_2 \times Z_3$ orbifolds yielding $E_6^{k=2}$ gauge groups. I will,
in particular, discuss in detail new issues arising in higher twisted
sectors of {\em non-prime\/} asymmetric orbifolds which were not worked out
in the original article on asymmetric orbifolds by Narain, Sarmadi and 
Vafa~\cite{NSV}. $SO(10)$ and $E_6$ models 
based on $Z_2 \times Z_4$ orbifolds are the subject of section~\ref{so10}. 
Here I show how to avoid the phase ambiguities of the type encountered
in section~\ref{E6}. I present my conclusions in section~\ref{summary}.

\section{Higher level string models}
\label{general}
There are three basic relations~\cite{GO} of relevance to higher level
string model building. The first one,
\be
\label{c}
   c = \frac{k\, {\rm dim\;} G}{k + \tilde{h}},
\ee
relates the central extension of the Kac-Moody algebra being proportional
to the level $k$ to its contribution to the conformal anomaly $c$, which 
in turn parametrizes the central extension of the Virasoro algebra. 
In Eq. (\ref{c}) ${\rm dim\;} G$ is the dimension of the gauge group $G$ 
and $\tilde{h}$ is the dual Coxeter number. The second relation,
\be
\label{h}
   h_R = \frac{C_R}{2(k + \tilde{h})},
\ee
gives the conformal dimension of a primary field transforming in representation
$R$ under $G$. $C_R$ is the quadratic Casimir invariant of $R$,
\be \label{casimir}
  \tr_{R_1} F^2 = {C_1 {\rm dim\;} R_1\over C_2 {\rm dim\;} R_2} \tr_{R_2} F^2,
\ee
where $F^2$ refers to the gauge field kinetic energy and
for the adjoint representation $A$ one has $C_A = 2\tilde{h}$. 
Finally, there is an inequality restricting the (unitary) representations $R$ 
in which primary fields can transform for given $k$,
\be
\label{k}
   k \geq \sum\limits_{i=1}^{{\rm rank\;} G} n_i m_i.
\ee 
Here $n_i$ are the Dynkin labels of $R$ and $m_i$ the co-marks of $G$.
The values for $\tilde{h}$ and $m_i$ can be found in Table~1 of 
Reference~\cite{Lewellen}. 

Applying Eq.~(\ref{c}) to the heterotic string ($c \leq 22$) one finds for the
exceptional groups $k \leq 4,3,2$ for $E_6, E_7, E_8$, respectively.
Eq.~(\ref{h}) then reveals that only the fundamental and adjoint 
representations of these groups can appear in the {\em massless\/} spectrum 
($h \leq 1$). Hence, the {\bf 351} and ${\bf 351^\prime}$ which are often used 
in $E_6$ model building are not permitted. 

$SO(4N+2)$ for $N \geq 2$ are candidate gauge groups for unified model
building, with each fermion generation in the non-selfconjugate, anomaly-free,
basic spinor representations $4^N$. However, Eq.~(\ref{c}) shows that for
$N > 3$ the level $k$ cannot be greater than 2. On the other hand, we find
for the basic spinor representation of $SO(2N)$,
\be 
{\rm at\; level\; 1}\;\;\; h_{2^{N-1}} = {N \over 8},
\ee
and
\be 
{\rm at\; level\; 2}\;\;\; h_{2^{N-1}} = {2N - 1 \over 16},
\ee
which both exceed 1 for $N \geq 9$, so that $SO(18)$ and bigger orthogonal
groups are ruled out from heterotic string model building. For $SO(10)$
and $SO(14)$ we find $k \leq 7$ and $k \leq 3$, respectively. Eq.~(\ref{k})
can now be used to establish that primary fields transforming in the {\bf 120} 
or {\bf 126} of $SO(10)$ can appear\footnote{At level 2, all two-index
(vector, spinor or mixed) representations, as well as arbitrary-index 
complete antisymmetric representations are allowed.} for $k \geq 2$. 
They play the analogous roles for the discussion of mass matrices like the 
{\bf 351} and ${\bf 351^\prime}$ of $E_6$, respectively.
However, as already noted in~\cite{AFIU}, condition~(\ref{h}) is often
much stronger. Indeed, the {\bf 120} can only appear in the massless spectrum
for $k \geq 3$, and the {\bf 126} (whose Clebsch-Gordon coefficients
could account for $m_s$ -- $m_\mu$ unification) for $k \geq 5$.

As for $SU(N)$ GUT models from the heterotic string, Eq.~(\ref{c}) yields 
$N \leq 12$ for level $k=2$. In general, no further restrictions arise, 
since the conformal dimension of the $M$-index complete antisymmetric
representation of $SU(N)$ is at level 1 given by
\be
h = \frac{M(N-M)}{2N} \;\;\; (< 1 \;\; {\rm for} \;\; M = 1,2).
\ee
Thus, the two-index antisymmetric and fundamental representations as used, 
for example, for minimal $SU(5)$ are allowed to appear in the massless 
spectrum for any $N$. The {\bf 45} of $SU(5)$, used to achieve more 
realistic fermion mass matrices, is allowed by both Eq.~(\ref{h}) and 
Eq.~(\ref{k}) to be massless for $k \geq 2$. The level two representations
{\bf 50} and {\bf 75}, employed in the missing partner mechanism~\cite{MNTY}, 
can appear in the massless spectrum for $k \geq 4$ and 
$k \geq 3$, respectively. Note, however, that for this mechanism explicit 
heavy mass terms are needed to keep the unwelcome states inside the {\bf 50} 
and {\bf 75} at the GUT scale. Therefore, it seems preferable to construct 
models where these representations have a mass which would be a fraction of the
Planck mass, using explicitly the fact that the unification and string scales
are close to each other. For example, one could make use of the small mass
increments in $Z_N$-orbifolds of high twist order $N$. There are two more
advantages for doing so: if the extra {\bf 50} and {\bf 75} representations
are too close to the GUT scale, then the gauge coupling would become 
strong below the Planck scale; on the other hand, if they are too close to the 
Planck scale the see-saw type triplet mass would be significantly below the GUT
scale, which would lead to too rapid proton decay~\cite{HMTY}. 

In fact, if one is interested in massive states, it is important to note
that Eq.~(\ref{k}) is a restriction on the unitary highest weight 
representations of the conformal field theory, i.e.\ it constraints 
the primary fields. However, given any primary field $\phi_R$, through 
secondary fields one will find all representations of $G$ in the same 
congruency class as $R$ at some mass level. By the same token, the Kac-Moody
currents being themselves descendents of the identity field can give rise
to massless adjoint representations already at level one. Examples are
gauge bosons, gauginos, as well as adjoint scalars in $N=2$ supermultiplets. 
However, these massless adjoints are always non-chiral~\cite{Lewellen}.

\section{Asymmetric Orbifolds}
\label{asymmetric}
In this section, I will describe how one can use the asymmetric orbifold
construction to manifestly arrive at level two models. I will not consider the 
possibility of quantized Wilson lines the inclusion of which is 
straightforward, but more tedious in practice. When there are no Wilson lines, 
the internal space part decouples from the gauge part and the Narain 
vectors~\cite{NSW} are simply given by
\be 
   P_{L/R} = {m \over 2} - b n \pm g n,
\ee
where $n$ and $m$ are integer valued $d$ dimensional winding and momentum 
vectors. The metric $g$ is normalized (using $\alpha^\prime$) such that the 
Narain scalar product is given by
\be
  P^2 \equiv P_L^2 - P_R^2 \equiv P_L g^{-1} P_L - P_R g^{-1} P_R.
\ee
With this convention, at a point of enhanced symmetry, $g$ is one quarter 
of the respective Cartan metric and the tensor $b$ is any antisymmetric 
counterpart of $g$. 

An orbifold twist $\th$ must leave the conformal dimensions $P_L^2/2$ and 
$P_R^2/2$ invariant. Hence, any twist has the form
\be
  P_{L/R} \rightarrow P_{L/R}^\prime = \th_{L/R}^\star P_{L/R},
\ee
with $\th_{L/R}^\star \equiv \th_{L/R}^{T^{-1}}$, and the two conditions, 
\be
\label{metcond}
  \th_{L/R}^t g \th_{L/R} = g,
\ee
must be satisfied. The transformed winding and momentum vectors are then 
straightforwardly obtained, and given by
\be
\ba{c}
n^\prime = a_{nn} n + a_{nm} m, \\
m^\prime = a_{mn} n + a_{mm} m, 
\ea
\ee
where
\be
\ba{l}
\label{twist}
   a_{nn} = {1\over 2} [(\th_L + \th_R) - (\th_L - \th_R) g^{-1} b], \vier 
   a_{nm} = {1\over 4} (\th_L - \th_R) g^{-1}, \vier
   a_{mn} = \tilde{\th}+ (\tls - \trs) g - b (\th_L - \th_R) g^{-1} b, \vier
   a_{mm} = {1\over 2} [(\tls + \trs) + b (\th_L - \th_R) g^{-1}],
\ea
\ee
and 
\be
\tilde{\th} = b (\th_L + \th_R) - (\tls + \trs) b.
\ee
The blocks $a_{ij}$ fill up a $2d\times 2d$ dimensional integer matrix,
\be
\label{bigt}
   \Theta = \left( \ba{cc} a_{nn} & a_{nm} \\ 
                        a_{mn} & a_{mm} \ea \right), 
\ee
which is the twist matrix acting in an Euclidean lattice with metric
\be
\label{bigg}
  G = \left( \ba{cc} 2 (g-b)g^{-1}(g+b) & bg^{-1} \\ 
                      -g^{-1}b & {1\over 2g} \ea \right).
\ee

Note, that for symmetric twists, $\th \equiv \th_L \equiv \th_R$, 
$n$ transforms homogeneously, 
\be
\label{ntrafo}
  n \rightarrow \th n,
\ee
whereas
\be
\label{mtrafo}
  m \rightarrow \th^\star m + \tilde{\th} n
\ee
receives a winding admixture whenever $2 [b \th - \th^\star b] \neq 0$.
One sees that $\th$ and $\tilde{\th}$ are integer valued
matrices and that $b$ can assume quantized values similar to Wilson lines. 
The $T$-dual twist~\cite{EJN},
\be 
   \hat{\theta}_{L,R}=\theta-2 (b\mp g)\tht=(g\mp b)\tha {1\over g\mp b},
\ee
is, however, asymmetric. Duality is here not a symmetry in moduli space, but
relates symmetric and asymmetric orbifolds. That symmetric and asymmetric
orbifolds are closely related is also evident from the fact that the K3
surface has both symmetric and asymmetric orbifold points~\cite{6D}. Thus,
an asymmetric orbifold can have a very clear geometrical interpretation.

I will now discuss a simple asymmetric orbifold at level 2. 
Although it is realized in $D=6$ uncompactified dimensions with simple 
supersymmetry, it is related to the non-supersymmetric $D=10$ 
heterotic string theory with gauge group $E_8^{k=2}$~\cite{KLT}. It will 
(sometimes with modifications of the compactification lattice) play the role
of the untwisted sector of the $D=4$ models to be discussed later.

It is a $Z_2$ orbifold in which the four dimensional internal space part on 
the bosonic side of the heterotic side remains untouched, $\th_L = {\bf 1}$, 
while $\th_R = - {\bf 1}$, and the two $E_8$ factors are interchanged. The 
compactification lattice can be uniquely determined by looking at the 
degeneracy in the twisted sector. In an even self-dual lattice 
the quantity~\cite{NSV}
\be
\label{deg}
  D =\sqrt{N^f_L \times N^f_R \over \det g_{inv}},
\ee
where $N^f_{L/R}$ are the numbers of left and right fixed points, 
is always an integer. $g_{inv}$ is the metric of the invariant
Narain sublattice. In the case at hand, the number of left fixed points 
(from the gauge part) is $2^8$. From the metric of the invariant gauge 
lattice (the {\em diagonal} $E_8$) we find
\be
   \det 2 g_{E_8} = 2^8,
\ee
as well, because the invariant $E_8$ vectors have double length squares.
Thus the degeneracy from the gauge part is $D_{gauge} = 1$. 
The number of right fixed points is $2^4$. Thus, the metric of the 
invariant space lattice must have determinant 1, 4 or 16. It is given by 
$4g$ which we want to be the Cartan matrix of a simply-laced Lie algebra. 
A look at the semi-simple Lie algebras with rank 4 reduces the choice to
$SU(2)^4$ or $SO(8)$. Both give rise to integral twists $\Theta$.
However, the $SU(2)^4$ lattice must be rejected, as it will become clear 
later that this lattice does not satisfy the level matching condition. On 
the other hand the $SO(8)$ lattice is adequate ($\det g_{SO(8)} = 4$, $D=2$).

The massless states in the untwisted sector are easily obtained: the 
supergravity and dilaton multiplets in six dimensions; the super Yang-Mills 
multiplets from the diagonal $E_8$ and the enhanced $SO(8)$; and one 
hypermultiplet transforming in the adjoint of $E_8$. As for the twisted sector,
we have already seen that $D = 2$, but since there are only 2 spinors per
fixed point and a hypermultiplet contains 4 fermionic states, we find an 
effective degeneracy $D_{\rm eff} = 1$. The vacuum energy for the left movers 
is 1/2. Upon world sheet modular transformations we have the lattice with 
metric 
$g_{inv}^{-1}$ at our disposal. Due to the self-duality of the $E_8$ lattice, 
the inverse of $2 g_{E_8}$ is up to an integral similarity transformation 
simply given by ${1\over 2} g_{E_8}$. Thus, for massless states we have 
to look for solutions of 
\be \label{p2} P^2_{E_8}/4 = 1/2, \ee 
which are just the roots. 
Combined with the 8 half-oscillator states they give rise to a twisted adjoint 
representation of $E_8$. Finally, for states with $P^2_{E_8} = 0$, we must 
consider the lattice with metric $g^{-1}_{SO(8)}$, the weight space of $SO(8)$.
Massless states are the ones with length squares equal to unity and correspond 
to the triality symmetric combination ${\bf 8}_v + {\bf 8}_s +{\bf 8}_c$. 

At this point level matching would break down, had we chosen the $SU(2)^4$
lattice. The states in the $SU(2)$ weight lattices have length squares 
corresponding to conformal dimensions $k^2/4$, with $k \in {\bf Z}$. 
Hence, states corresponding to odd $k$ do not match with states from the 
right hand side which are half integer spaced.

In total we have matter transforming under $E_8 \times SO(8)$ like
$$ 2 ({\bf 248},{\bf 1}) + ({\bf 1}, {\bf 8}_v + {\bf 8}_s +{\bf 8}_c). $$
Note that 
\be
   N_H - N_V = 244,
\ee
where $N_H$ and $N_V$ are the numbers of hypermultiplets and vector multiplets,
respectively, as is required in six dimensional supergravity with precisely
one tensor multiplet (the dilaton multiplet)
for the cancellation of gravitational anomalies. Also, as usual in $D=6$, 
cancellation of gauge and mixed gauge/gravitational anomalies constitutes
a highly non-trivial check. The anomaly has to factorize so that it can be 
canceled by the Green-Schwarz mechanism~\cite{GS}. Here it does, and there is 
a new feature at higher level. The anomaly polynomial is given by
\be
\label{anomaly}
 i(2 \pi)^3 I = -{1\over 16}
          [\tr R^2 - {1\over 6} \Tr F^2_{SO(8)} - {1\over 15}\Tr F^2_{E_8}]
   \times [\tr R^2 + {1\over 4} \Tr F^2_{SO(8)} - {1\over 10}\Tr F^2_{E_8}], 
\ee
where traces in the adjoint representations of the gauge groups are used.
Note, that then the coefficients in the first factor are simply given by 
$k/\tilde{h}$. One can use this fact to show that in $D=6$ there can only be
three possibilities for $E_8$: (1) no adjoint representation corresponding to 
$k=1$ like in the case of compactifying the heterotic string on $K_3$;
(2) one adjoint corresponding to $k=0$, i.e. the {\bf 248} must be part
of an $N=2$ gauge multiplet; and (3) two adjoints as in the $k=2$ case
just discussed. A larger number of {\bf 248} representations, would lead 
to irrational coefficients in the anomaly polynomial. 

For compactification to $D=4$, $N=2$, the above model can now be used by 
either attaching a torus, $T_2$, or by changing the lattice to e.g.\
the $SO(12)$ or the $E_6$ lattice. The breaking to $N=1$ goes along with a 
simultaneous breaking of $E_8$. One can act in each $E_8$ with a twist or
shift of order $N$, but it must be the same action in both $E_8$ factors.
This defines an auxiliary $Z_N$ orbifold in its own right which would give
rise to a gauge group $G_8 \times G_8 \times G_6$. $G_8$ is the gauge group
which (upon permutation of the two sets of $E_8$ gauge coordinates)
will be promoted to level 2 and $G_6$ is the enhanced gauge group arising
from the space part.

To summarize, the class of models described in detail below, are 
$Z_2 \times Z_N$ orbifolds, where $Z_2$ refers to the level 2, $N=2$ models 
above, and $Z_N$ breaks one supersymmetry and $E_8$ to $G_8$. In the
simplest case of a $Z_2 \times Z_2$ orbifold one can obtain only the
non-chiral groups $G_8 = E_7 \times SU(2)$ and $SO(16)$. On the other hand,
$Z_2 \times Z_3$ and $Z_2 \times Z_4$ orbifolds yield many chiral possibilities
as shown in Table~\ref{groups}, and
\begin{table}[b]
\centering
\begin{tabular}{|c|c|c|c|} \hline
 $m$ & $Z_2 \times Z_2$   & $Z_2 \times Z_3$    & $Z_2 \times Z_4$ \\ \hline

  1  & $E_7 \times SU(2)$ & $E_7\times U(1)$    & $E_7\times U(1)$ \\
  2  & $SO(16)$           & $SO(14)\times U(1)$ & $SO(14)\times U(1)$ \\
  3  &               & $E_6\times SU(3)$   & $E_6\times SU(2)\times U(1)$ \\
  4  &               & $SU(9)$             & $SU(8)\times U(1)$ \\
  5  &               &                     & $SO(12)\times SU(2)\times U(1)$ \\
  6  &               &                     & $SO(10)\times SU(4)$ \\
  7  &               &                     & $SU(7)\times SU(2)\times U(1)$ \\
\hline
\end{tabular}
\caption{Possible groups $G_8$ at level $k=2$ from $Z_2 \times Z_N$ orbifolds. 
The groups are listed according to the twist order $N$ and an integer $m$. 
The gauge contribution to the vacuum energy of the first twisted sector is 
given by $E_{vac}^{gauge} = 2 m/N^2$.}
\label{groups}
\end{table}
$Z_2 \times Z_6$ orbifolds include $SU(5)\times SU(4)\times U(1)$.
In the next section, I will exploit the most interesting $Z_2 \times Z_3$
case, namely $E_6\times SU(3)$ models. Section~\ref{so10} 
focuses on $Z_2 \times Z_4$ orbifolds with gauge groups $SO(10)\times SU(4)$
and $E_6\times SU(2) \times U(1)$.

\section{$Z_2\times Z_3$ orbifolds with $E_6$ gauge group at level 2}
\label{E6}

The $Z_3$ action of the $Z_2 \times Z_3$ orbifold in the gauge part is the 
same as in the case of the standard ``$Z$ manifold''~\cite{DHVW}, only that 
here both $E_8$ factors are twisted\footnote{In Reference~\cite{DMW} this is 
called symmetric embedding.}. It is a peculiarity of prime orbifolds, that they
lead to modular invariant partition functions when one uses standard embeddings
without twisting the left space part. That opens up the two possibilities of 
twisting all left internal coordinates by a $Z_3$ rotation, or alternatively, 
leaving all of them untouched.

On the right hand side we have two choices of supersymmetric $Z_3$ twists:
$Z_3$ could act like in the case of the $Z$ manifold by rotating
all three pairs of right handed internal coordinates, or it could rotate
just two pairs, in which case it leads by itself to $N=2$ supersymmetry and 
I will refer to it as $Z_3^\prime$. Let the $Z_2$ action take place in the 
first two complex coordinates of the right hand side and the $Z_3^\prime$
action in the last two. Then, $Z_2 \times Z_3$ corresponds to the usual
$Z_6$ orbifold and $Z_2 \times Z_3^\prime$ to $Z_6^\prime$. In summary, we have
four possibilities to choose the internal {\em twist eigenvalue structure}:
\be
\ba{llcccl}
     {\rm A} & ( & Z_3     &, & Z_6        &), \\
     {\rm B} & ( & {\bf 1} &, & Z_6        &), \\
     {\rm C} & ( & Z_3     &, & Z_6^\prime &), \\
     {\rm D} & ( & {\bf 1} &, & Z_6^\prime &). 
\ea
\ee
The next step consists of specification of the lattices. 

\underline{Cases A through C:} In these cases there is at least one $Z_3$ 
involved (as opposed to $Z_3^\prime$). That means one has to look for groups 
possessing an $SU(3)^3$ subgroup. The two possibilities are $E_6$ and $SU(3)^3$
itself. However, the $SU(3)^3$ root lattice must be rejected, because a $Z_6$ 
twist cannot act asymmetrically in an $SU(3)$ lattice.

On the other hand, a consistent twist acting in the $E_6$ lattice can be 
constructed. It is convenient to use the $SU(3)^3$ basis of $E_6$.
Define the simple roots of one $SU(3)$ by
\be e_1 = (\sqrt{2},0), \;\;\; e_2 = (-{1\over \sqrt{2}},\sqrt{3\over2}). \ee
The Cartan metric is
\be
  g_{SU(3)} = \left( \ba{rr} 2 & -1 \\ -1 & 2 \ea \right),
\ee
and the fundamental weights are given by
\be \tilde{e}_1 = ({1\over \sqrt{2}},{1\over \sqrt{6}}), \;\;\; 
   \tilde{e}_2 = ( 0               ,\sqrt{2\over3}). \ee
Note, that
\be
\ba{c}
           e_1 +         e_2 = \tilde{e}_1 + \tilde{e}_2, \\
   \tilde{e}_2 - \tilde{e}_1 = \tilde{e}_1 - e_1,
\ea
\ee
which is useful for constructing the twists.
Distinguish between the $SU(3)$ factors by unprimed, primed and doubly
primed symbols. Then, a basis of $E_6$ is given by
\be
\ba{ll}
   f_1 = (e_1,0,0), & f_2 = (e_2,0,0), \\
   f_3 = (0, e_1^\prime,0), & f_4 = (0, e_2^\prime,0), \\
   f_5 = (\tilde{e}_1, \tilde{e}_1^\prime, \tilde{e}_1^{\prime\prime}), &
   f_6 = (\tilde{e}_2, \tilde{e}_2^\prime, \tilde{e}_2^{\prime\prime}),
\ea
\ee
corresponding to the metric
\be
   g_{E_6} = \left( \ba{rrr} 
   g_{SU(3)} & 0 & {\bf 1} \\ 0 & g_{SU(3)} & {\bf 1} \\ {\bf 1} & {\bf 1} & 
   3 g_{SU(3)}^{-1} \ea \right),
\ee
with inverse
\be
   g_{E_6}^{-1} = \left( \ba{rrr} 
   2 g_{SU(3)}^{-1} &   g_{SU(3)}^{-1} & -{\bf 1} \\
     g_{SU(3)}^{-1} & 2 g_{SU(3)}^{-1} & -{\bf 1} \\ 
     -{\bf 1}       &   -{\bf 1}       & g_{SU(3)} \ea \right).
\ee
If we denote a $Z_3$ twist in one $SU(3)$ by 
\be \th = \left( \ba{rr} 0 & -1 \\ 1 & -1 \ea \right), \ee
we can write for the six dimensional $Z_3$ twist matrix
\be \th_3 = {\rm diag} (\th,\th,\th^\star). \ee
If we further define
\be \Delta \equiv \left( \ba{rr} 1 & 1 \\ 0 & 1 \ea \right), \ee
we have for $Z_3^\prime$
\be \th_3^\prime = \left( \ba{rrr} 
  \th & 0 & -\Delta \\ 0 & \th & -\Delta \\ 0 & 0 & {\bf 1} \ea \right). 
\ee
Finally, we have to specify in which way the $Z_2$ acts in our lattice.
We choose it in such a way that it permutes the first two $SU(3)$ factors
in addition to negating all vectors (in order to get the correct number
of eigenvalues $-1$),
\be
 \th_2 = \left( \ba{rrr} 
     0 & -{\bf 1} & 0 \\ -{\bf 1} & 0 & 0 \\ 0 & 0 & -{\bf 1} \ea \right).
\ee
It can be checked that $g_{E_6}$ (in the sense of Eq.~(\ref{metcond})), 
$\th_3$, $\th_3^\prime$ and $\th_2$ all
mutually commute. Now we can simply define $\th_6 = \th_2 \th_3$ and
$\th_6^\prime = \th_2 \th_3^\prime$. For the antisymmetric tensor $b$ we choose
\be
   b_{E_6} = \left( \ba{rrr} 
   \sigma & 0 & {\bf 1} \\ 0 & \sigma & {\bf 1} \\ -{\bf 1} & -{\bf 1} & 
   \sigma \ea \right),
\ee
with 
\be \sigma \equiv \left( \ba{rr} 0 & 1 \\ -1 & 0 \ea \right), \ee
but the distribution of signs plays no role. One still has to show
that the twists defined this way lead to an integer matrix $\Theta$
when inserted into the expressions (\ref{twist}). This turns out to be
true for all cases A through D.

\ul{Case D} does not involve the $Z_3$ (only $Z_3^\prime$) twist, and
one can try to find more lattices than just the root lattice of $E_6$.
Indeed, since case D possesses 12 right fixed points, the metric of the 
invariant sublattice could have a determinant of either 3 or 12. The 
$E_6$-lattice discussed before corresponds to the former case since 
$\det g_{E_6} = 3$. A lattice with determinant 12 is the root lattice 
of $SO(8) \times SU(3)$. We define $Z_3^\prime$ such that it acts in the 
explicit $SU(3)$ factor and in an $SU(3)$ subgroup of $SO(8)$. The $Z_2$ 
acts by negating the $SO(8)$ roots. Again, the integer condition following 
from (\ref{twist}) can be checked to be satisfied.

I will refer to the orbifolds defined by the eigenvalue structures A through D 
acting in the $E_6$ lattice as models A through D and model E will be the one 
realized in $SO(8) \times SU(3)$. The resulting spectra are displayed in 
Table~\ref{spectra}. In the following, I discuss in some detail the 
spectrum calculation for model A. In the second and third twisted sector, we 
will find the phase ambiguities alluded to in the introduction. In all 
models A through E the phases can be fixed by requiring CPT invariance and 
cancellation of anomalies. Clearly, this is not a satisfactory state of 
affairs, and in section~\ref{so10}, I will introduce a systematic way to 
compute such phases.
\begin{table}[t]
\begin {tabular}{|c|c|c|c|c|c|} \hline 

 & A & B & C & D & E \\ \hline \hline $G_6$ &
$SU(3)^3$ &
$E_6$ &
$SU(3)^3$ &
$E_6$ &
$SU(3) \times SO(8)$ \\ \hline & & & 

$({\bf 78},{\bf 1},{\bf 1},{\bf 1},{\bf 1})$ &
$({\bf 78},{\bf 1},{\bf 1})$ &
$({\bf 78},{\bf 1},{\bf 1},{\bf 1})$ \\ & & &

$({\bf 1},{\bf 8},{\bf 1},{\bf 1},{\bf 1})$ &
$({\bf 1},{\bf 8},{\bf 1})$ &
$({\bf 1},{\bf 8},{\bf 1},{\bf 1})$ \\ U & & &

$({\bf 27},{\bf 3},{\bf 1},{\bf 1},{\bf 1})$ &
$({\bf 27},{\bf 3},{\bf 1})$ &
$({\bf 27},{\bf 3},{\bf 1},{\bf 1})$ \\ &

$ 3 (\bar{\bf 27},\bar{\bf 3},{\bf 1},{\bf 1},{\bf 1})$ &
$ 3 (\bar{\bf 27},\bar{\bf 3},{\bf 1})$ &
$   (\bar{\bf 27},\bar{\bf 3},{\bf 1},{\bf 1},{\bf 1})$ &
$   (\bar{\bf 27},\bar{\bf 3},{\bf 1})$ &
$   (\bar{\bf 27},\bar{\bf 3},{\bf 1},{\bf 1})$ \\ &

$({\bf 1},{\bf 1},{\bf 3},{\bf 3},{\bf 3})$ & &
$({\bf 1},{\bf 1},\bar{\bf 3},\bar{\bf 3},\bar{\bf 3})$ & & \\ \hline & &

$   ({\bf 27},\bar{\bf 3},{\bf 1})$ & &
$ 2 ({\bf 27},\bar{\bf 3},{\bf 1})$ &
$   ({\bf 27},\bar{\bf 3},{\bf 1},{\bf 1})$ \\ T1 &

$   ({\bf 1},{\bf 3},\underline{{\bf 3},{\bf 1},{\bf 1}})$ & &
$ 2 ({\bf 1},{\bf 3},\underline{{\bf 3},{\bf 1},{\bf 1}})$ & &
$   ({\bf 1},{\bf 3},{\bf 3} + \bar{\bf 3},{\bf 1})$ \\ & & & & &
                                                     \\ \hline & & & &

$({\bf 27},\bar{\bf 3},{\bf 1})$ &
$({\bf 27},\bar{\bf 3},{\bf 1},{\bf 1})$ \\ & & 

$ 3 (\bar{\bf 27},{\bf 3},{\bf 1})$ & &
$   (\bar{\bf 27},{\bf 3},{\bf 1})$ & 
$   (\bar{\bf 27},{\bf 3},{\bf 1},{\bf 1})$ \\ T2 &

$({\bf 1},{\bf 6},\underline{\bar{\bf 3},{\bf 1},{\bf 1}})$ & & 
$({\bf 1},\bar{\bf 6},\underline{{\bf 3},{\bf 1},{\bf 1}})$ & &
$({\bf 1},{\bf 6},{\bf 3}+\bar{\bf 3},{\bf 1})$ \\ &

$ 2 ({\bf 1},\bar{\bf 3},\underline{\bar{\bf 3},{\bf 1},{\bf 1}})$ & &
$   ({\bf 1},\bar{\bf 3},\underline{\bar{\bf 3},{\bf 1},{\bf 1}})$ & &
$   ({\bf 1},{\bf 3},{\bf 3}+\bar{\bf 3}, {\bf 1})$ \\ & & & & &
                                                    \\ \hline &

$({\bf 78},{\bf 1},{\bf 1},{\bf 1},{\bf 1})$ &
$({\bf 78},{\bf 1},{\bf 1})$ & & &
$({\bf 78},{\bf 1},{\bf 1},{\bf 1})$ \\ & 

$({\bf 1},{\bf 8},{\bf 1},{\bf 1},{\bf 1})$ &
$({\bf 1},{\bf 8},{\bf 1})$ & & &
$({\bf 1},{\bf 8},{\bf 1},{\bf 1})$ \\ T3 &

$   ({\bf 27},{\bf 3},{\bf 1},{\bf 1},{\bf 1})$ &
$   ({\bf 27},{\bf 3},{\bf 1})$ &
$ 2 ({\bf 27},{\bf 3},{\bf 1},{\bf 1},{\bf 1})$ &
$ 2 ({\bf 27},{\bf 3},{\bf 1})$ & \\ & & & & &

$(\bar{\bf 27},\bar{\bf 3},{\bf 1},{\bf 1})$ \\ & & & & &
$({\bf 1},{\bf 1},{\bf 1},{\bf 8}_v+{\bf 8}_s+{\bf 8}_c)$ \\ \hline 
\end{tabular}
\caption{Models from asymmetric $Z_2 \times Z_3$ orbifold with 
         gauge group $[E_6 \times SU(3)]^{k=2} \times G_6^{k=1}$.
         U denotes the untwisted sector, while T1, T2 and T3 are the 
         twised sectors.}
\label{spectra}
\end{table}

\subsubsection*{Untwisted sector}

For the NSR-fermions I will use the shift description, i.e.\ I use
bosonized world sheet fermions and act with shift vectors in the vector and 
spinor concruency classes of the $SO(8)$ lattice. In the explicit discussion, 
I restrict myself to positive helicity spinor states (last entry $= + 1/2$), 
as the remaining states are just the CPT and supersymmetry partners. 
The relevant shift vectors $v$ corresponding to the $Z_6$ and $Z_6^\prime$ 
twists discussed before will be taken to be
\be
\ba{c}
   v_6        = (+{1\over 6}, +{1\over 6}, -{1\over 3}, 0), \vier
   v_6^\prime = (+{1\over 6}, +{1\over 3}, -{1\over 2}, 0), \vier
   v_3        = (+{1\over 3}, +{1\over 3}, -{2\over 3}, 0), 
\ea
\ee
where for comparison the shift vector for the standard $Z_3$ orbifold is
also shown. The positive helicity ground states $h$ with their shift phases 
$e^{2 \pi i h v}$ for the three cases are ($\alpha = e^{2 \pi i/6}$)
\be
\ba{c|c|c|c}
     \th_R      &  Z_6     &   Z_6^\prime &   Z_3     \\ \hline
(+{1\over 2},+{1\over 2},+{1\over 2},+{1\over 2})&  1  &  1  &  1  \vier
(+{1\over 2},-{1\over 2},-{1\over 2},+{1\over 2})& \alpha & \alpha & \alpha^2
\vier
(-{1\over 2},+{1\over 2},-{1\over 2},+{1\over 2})& \alpha & \alpha^2&\alpha^2
\vier
(-{1\over 2},-{1\over 2},+{1\over 2},+{1\over 2})& \alpha^4 &  -1  & \alpha^2 
\ea
\ee 
The gauge shift vector is defined as
\be
   V_3^{gauge} = (v_3,0^4;v_3,0^4).
\ee

Consider the {\bf 128} spinor representation of $SO(16) \subset E_8$, 
which decomposes into 8 groups of {\bf 16} (${\bf \bar{16}}$) of $SO(10)$. 
Labeling these groups by their first three entries one finds the following 
twist phases and gauge transformation properties under $E_6 \times SU(3)$: 
\be
\label{gaugetrafo}
\ba{c|c|c|c}
                                      & S & A & \\ \hline 
(+{1\over 2},+{1\over 2},+{1\over 2}) & 1 & -1 & \vier
(-{1\over 2},-{1\over 2},-{1\over 2}) & 1 & -1 & 
({\bf 78},{\bf 1}) + ({\bf 1},{\bf 8}) \vier \hline
(+{1\over 2},-{1\over 2},-{1\over 2}) & \alpha^2 & \alpha^5 &  \vier
(-{1\over 2},+{1\over 2},-{1\over 2}) & \alpha^2 &\alpha^5&({\bf 27},{\bf 3})
\vier
(-{1\over 2},-{1\over 2},+{1\over 2}) & \alpha^2 & \alpha^5 &  \vier \hline
(-{1\over 2},+{1\over 2},+{1\over 2}) & \alpha^4 &  \alpha  &  \vier
(+{1\over 2},-{1\over 2},+{1\over 2}) & \alpha^4 &  \alpha  & 
                                              (\bar{\bf 27},\bar{\bf 3}) \vier
(+{1\over 2},+{1\over 2},-{1\over 2}) & \alpha^4 &  \alpha  &
\ea
\ee
In this table $S$ and $A$ refer to the symmetric and antisymmetric linear
combinations of the two $E_8 \times E_8$ vectors. The twist invariant states
will yield the gauge bosons. We can see that we have untwisted adjoint matter, 
iff there is a helicity vector with twist phase $-1$. This is not the case
for model A where we find fields transforming like
$$ 3 (\bar{\bf 27},\bar{\bf 3},{\bf 1}), $$
under the unbroken gauge group $[E_6\times SU(3)]^{k=2} \times G_6^{k=1}$. For 
model A, $G_6 = SU(3)^3$ since only 24 orbits of the 72 roots of $E_6$ are 
twist invariant. 48 orbits and the 6 oscillators transform under the twist.
Therefore there are additional untwisted matter fields transforming like
$$ ({\bf 1},{{\bf 1},\bf 3},{\bf 3},{\bf 3}). $$ 
In this class of models untwisted
adjoint matter appears precisely when the untwisted sector is non-chiral.

For comparison I have chosen a convention in which the ordinary $Z_3$ orbifold 
at the point of maximally enhanced gauge symmetry has the spectrum
$$ \ba{r} 
3 ({\bf 1},{\bf 27},\underline{{\bf 3},{\bf 1},{\bf 1},{\bf 1}}), \vier
3({\bf 1},{\bf 1},\underline{\bar{\bf 3},\bar{\bf 3},\bar{\bf 3},{\bf 1}}),
\ea $$
under $E_8 \times E_6 \times SU(3)^4$, where underlining means to take all 
permutations. 

\subsubsection*{First twisted sector}

There is only one massless spinor in this sector, namely
\be
   p        = (-{1\over 3}, -{1\over 3}, +{1\over 6}, +{1\over 2}), 
\ee
having positive helicity. 
The degeneracy from the space part is $D=9$, and the corresponding fixed 
points are charged under the enhanced gauge group. The degeneracy from the 
gauge part is 1 since the number of gauge fixed points cancels the volume 
factor of the invariant gauge lattice. The latter is the {\em diagonal} $E_8$, 
with a shift vector $\tilde{V}^{gauge}_3$ acting in it. It is important to 
realize that $\tilde{V}^{gauge}_3$ is given by {\em twice\/} $V_3^{gauge}$ 
truncated to one $E_8$,
\be
   \tilde{V}^{gauge}_3 = 2 (v_3,0^4).
\ee
The vacuum energy from the gauge (space) part is 1/2 (1/3) so that we have to 
look for states satisfying
\be
   {(P_{E_8} + \tilde{V}^{gauge}_3)^2\over 4} = {1\over 6},
\ee
corresponding to a ({\bf 1},{\bf 3}). In order to determine the transformation
of the 9 fixed points, it suffices to compare with the spectrum of the 
$Z_3$ orbifold. In its simply twisted sector the {\bf 27} come together
with triplets of $SU(3)$ and since the helicities are positive in either case,
we have matter transforming like
$$({\bf 1},{\bf 3},\underline{{\bf 3},{\bf 1},{\bf 1}}).$$

\subsubsection*{Second twisted sector}

The degeneracy factor for model A is before projecting onto $Z_2$ invariant 
states easily seen to be given by $D=27$. However, in the case of non-trivial
invariant lattices, it may be less straightforward to find the degeneracy 
factor, and I will now shortly describe how to find it for model E. 
One notes first that from the $SU(3)$ factor we have $D=1$ since the
contribution of the three right-chiral fixed points is canceled by the 
invariant left-chiral $SU(3)$ root lattice. 
We have one more $Z_3$ acting in a subgroup of $SO(8)$, the rest 
being unrotated. We have to find the determinant of this invariant Narain 
sublattice. Combining the left and right parts of the Narain lattice, 
one finds an $SO(8) \times SO(8)$ sublattice. The metric of this lattice has 
determinant 16, while a self-dual lattice must have determinant one.
There must be extra states having entries simultaneously in the left and the 
right part of the Narain lattice. Integrality of self-dual lattices
requires them to be weight vectors. Furthermore, if
they correspond to Kaluza-Klein states, they are left-right symmetric. Thus,
one has to include the congruency class $({\bf 8_v},{\bf 8_v})$ which
after passing to a Euclidean metric enlarges $SO(8) \times SO(8)$ to 
$SO(16)$. Similarly, one has to add the classes $({\bf 8_s},{\bf 8_s})$ and 
$({\bf 8_c},{\bf 8_c})$, which combined transform as a {\bf 128} of $SO(16)$ 
so that one finally arrives at the $Spin(16)/Z_2 \equiv E_8$ lattice. In other
words, for a compactification on a lattice with metric $g = 1/4 g_{SO(8)}$ and
where $b$ is its antisymmetric counterpart, one finds from Eq.~(\ref{bigg}),
$G=g_{E_8}$. Now a $Z_3$ twist acting in an $SU(3)$ subgroup of $E_8$ leaves 
an $E_6$ root lattice invariant, the metric of which has determinant 3, 
cancelling the contribution from the three fixed points and there is 
an overall degeneracy of $D=1$. 

Similarly, the Narain lattice of model D contains the congruency classes 
$({\bf 78}, {\bf 1}) + ({\bf 1},{\bf 78}) + 
        ({\bf 27},{\bf 27}) + (\ol{\bf 27},\ol{\bf 27})$ of $E_6\times E_6$. 
The $Z_3$ acts in an $SU(3)\times SU(3)$ subgroup of the right-moving $E_6$, 
leaving fixed the root lattice of $E_6\times SU(3)$. Again, the 9 right fixed 
points cancel against its volume factor, $\det g_{E_6} \det g_{SU(3)} = 9$, 
yielding a degeneracy of $D=1$. 
These results can be checked explicitly, by solving the equation 
$\Theta N = N$, with $\Theta$ from Eq.~(\ref{bigt}) and $N^T = (n^T,m^T)$.

The massless spinor in this sector is given by
\be
   p        = (-{1\over 6}, -{1\over 6}, -{1\over 6}, +{1\over 2}).
\ee
As for the gauge part, for model A we have to consider solutions to
\be 
\label{sol3}
     {(P_{E_8} + 2 V^{gauge}_3)^2\over 2} = {2\over 3},
\ee
which correspond to 
\be \label{x33}   ({\bf 1},{\bf 3},{\bf 1},{\bf 3}). \ee
Now one has to project onto $Z_2$ invariant states. As mentioned there are 
27 fixed points under $Z_3$. Four complex {\em chiral\/} dimensions are purely
$Z_3$ rotated so that there is no $Z_2$ phase, but two complex dimensions 
behave non-trivially. For each such complex dimension the origin is fixed and 
there is the symmetric and the antisymmetric combination of the remaining 
$Z_3$ fixed points. Hence, we have two invariant (symmetric) and one 
antisymmetric combination, and I will denote this by $2^s + 1^a$. Combining 
everything yields
\be \label{deg1} D = \sqrt{3^4 (2^s + 1^a)^2} = \sqrt{3^4 (1^s + 2^a)^2} =
          9 (2^s + 1^a)\;\; {\rm or}\;\; 9 (1^s + 2^a), \ee
and one finds an ambiguity due to the presence of the square root.
Unfortunately, it is impossible to resolve this ambiguity in this
framework. In section~\ref{so10}, I will introduce a different, yet
equivalent method to describe these sort of models. It will use a shift
description also for the internal space part, so that all phases can be 
fixed unambiguously. Of course, one may also fix the phases by requiring
the model to be free of gauge anomalies. In any case it turns out that
the correct choice is the latter option in Eq. (\ref{deg1}) and to take the 
symmetric combination of (\ref{x33}), 
\be \label{states} {\bf 3} \times {\bf 3} = {\bf 6}^s + \bar{\bf 3}^a, \ee
with multiplicity 9 and the antisymmetric one with 
multiplicity 18. These states transform in addition under the
enhanced $SU(3)^3$. We conclude there is matter transforming like
$$   ({\bf 1},{\bf 6},\underline{\bar{\bf 3},{\bf 1},{\bf 1}}), $$
$$ 2 ({\bf 1},\bar{\bf 3},\underline{\bar{\bf 3},{\bf 1},{\bf 1}}). $$

A similar ambiguity appears in model C. Here are two opposite
helicities $p^\prime_\pm$ transforming with a relative sign under $Z_2$, 
$e^{6\pi i p^\prime_\pm v_6^\prime} = \alpha^5,\alpha^2$. Obviously, there
must be extra overall phases combining to $\pm \alpha$ in order to reduce
above phases to $\pm 1$. Again we cannot determine them within in the
present framework. Although it is clear that one has to take both the
symmetric and antisymmetric combinations appearing in the product 
(\ref{states}), the overall phase is important to determine chiralities.
Here it turns out that one
has to take the antisymmetric combination for the positive helicity states 
(yielding antitriplets) and the symmetric combination for negative helicity 
states (producing antisextets). A similar situation as in model C
occurs also in model E, but models B and D happen to be free of any
phase ambiguities.

\subsubsection*{Third twisted sector}

In this sector there are two massless spinors with opposite helicities,
\be
   p_\pm        = ( 0 , \;\;0 \mp {1\over 2}, \pm {1\over 2}).
\ee
The degeneracies for all models is $D=2$, as was the case for the 
$N=2$ model discussed at the end of section~\ref{asymmetric}.
This is evident for model E, but the fact that the asymmetric $Z_2$ action 
in the $E_6$ lattice indeed yields an invariant lattice of determinant
4 must be checked explicitly. From the gauge part one has the solutions 
of Eq.~(\ref{p2}) plus the eight half-integer oscillators at ones disposal.

Now one has to project onto $Z_3$ invariant states.
One obtains for the positive (negative) helicity vector
$e^{4\pi i p_\pm v_6} = \alpha^2 (\alpha^4)$. 
We are looking for $Z_2$ fixed points which are not fixed under $Z_3$. 
Consider first the complex dimension being $Z_6$ twisted. 
There is the twist invariant origin and the 
three non-trivial $Z_2$ fixed points transform as a triplet under $Z_3$. 
Hence, in a notation similar to the one of the previous sector one can write
\be
\label{deg2}
  D = \sqrt{2^s + 1^{\alpha^2} + 1^{\alpha^4}} 
    = \sqrt{(1^{\alpha^2} + 1^{\alpha^4})(1^{\alpha^2} + 1^{\alpha^4})}
    = 1^{\alpha^2} + 1^{\alpha^4}.
\ee
Fortunately, here the square root can be taken unambiguously, and the
relative phase corresponding to the twofold degeneracy is $\alpha^2$.
The other $Z_6$ action arises through the permutation of the remaining
two $SU(3)$ subgroups. This gives four chiral fixed points, but as a rule,
fixed points from permutation (sub-) orbifolds do not affect the degeneracy 
as their contribution is canceled against the volume factor of the invariant 
lattice. Thus from here cannot arise any {\em relative\/} phase.

Finally, we have to clarify whether there are some ambiguous {\em overall\/}
phases in this sector, as was the case in the second twisted sector. 
The answer is that there are none, because any possible ambiguity can be 
resolved in the following way: untwisted and order two twisted sectors must by 
themselves be CPT invariant. That means that all phases must come in 
complex conjugate pairs. As shown, they already have this property so that
there cannot be extra overall phases\footnote{In general, there could still
be an overall sign. For the present case, however, that would result in
phases which are not third roots of unity, which are the only possible ones
in a $Z_3$ projection.}.

Combining, finally, the internal phases with the NSR-phases we find that the 
positive helicity states are associated with $Z_3$ phases 1 and $\alpha^4$. 
In the gauge part projections have to made using $\tilde{V}^{gauge}_3 
= 2 (v_3,0^4)$. The factor 2 in the shift vector means that the obtained 
twist phases must match the {\em squares\/} of the phases shown in column 
$S$ of table (\ref{gaugetrafo}). 
Hence, we find matter transforming like
$$ ({\bf 78},{\bf 1},{\bf 1},{\bf 1},{\bf 1}) + 
   ({\bf  1},{\bf 8},{\bf 1},{\bf 1},{\bf 1}) + 
   ({\bf 27},{\bf 3},{\bf 1},{\bf 1},{\bf 1}). $$
In contrast to the untwisted sector, from this sector can arise adjoint 
representations even if it is chiral.

The complete spectra of all models A through E are shown in 
Table~\ref{spectra}.

\section{$Z_2\times Z_4$ orbifolds with $E_6$ and $SO(10)$ at level 2}
\label{so10}

In the course of the calculation in the last section, we encountered sign 
ambiguities which could not be resolved using standard asymmetric orbifold 
technology. In these cases, it was possible to fix the signs by simply 
requiring cancellation of anomalies, or by using other consistency arguments. 
In general, however, this information is
insufficient. Moreover, in quite involved calculations one would rather 
reserve anomaly factorizations and cancellations as cross check.

The easiest way to resolve these ambiguities is to avoid twist
rotations and to use instead space shifts leading to equivalent 
models\footnote{The example of space shifting the $Z_3\times Z_3$ orbifold 
was given in Reference~\cite{CK}.}. Then all phases can be determined in a
straightforward way, as will be worked out below in an example. I will
present this example in considerable detail since there are many non-trivial
phases arising in asymmetric orbifolds, which are unheard of in the symmetric
case where most of them cancel between left and right movers. 

Before launching the sample calculation, I first define eight models in
the $Z_2\times Z_4$ orbifold class. Actually, each of these models
grants the option of an extra discrete torsion sign~\cite{Vafa}
in the twist sector projectors. I will refer to (not) including this extra 
sign as negative (positive) discrete torsion. Models I through IV have level 2 
gauge group $SO(10)\times SU(4)$ while models V through VIII have 
$E_6 \times SU(2) \times U(1)$. The level $k$ of a $U(1)$ is meaningful: 
it describes an absolute normalization in which only certain charges may 
appear for massless states; moreover, the gauge transformation of
the antisymmetric tensor field (the duality transformed axion) is
proportional to $k$ which is important for the demonstration that a
potential $U(1)$ anomaly is canceled by the Green-Schwarz mechanism. 

Consider first the $Z_4$ suborbifold of models I through IV. When twisting all 
gauge coordinates of an $E_8$ lattice by $Z_4$, there are 60 twist invariant 
orbits of the 240 roots corresponding to the gauge group $SO(10)\times SU(4)$. 
Twisting both $E_8$ lattices in this way yields $E_{vac}^{gauge} = 3/4$. 
An equivalent gauge shift can be chosen to have the form
\be
   V_4^{gauge} = (V_4;V_4),
\ee
with
\be
\ba{c}
   V_4 = (+{1\over 2}, +{1\over 2}, +{1\over 2},0^5).
\ea
\ee
Given $E_{vac}^{gauge} = 3/4$, we have the options of twisting 
the left-handed space part by a four dimensional $Z_2$ reflection
(denoted $Z_2^\prime$ in the following), or 
not touching it at all. On the right hand side we have two choices of 
supersymmetric $Z_4$ twists which are discussed below. They will be called
$Z_4$ and $Z_4^\prime$. All these twists can be realized on $SO(12)$
lattices. In summary, we have four possibilities to choose the 
twist eigenvalue structure:
\be
\ba{lrcccl}
     {\rm I} & (& Z_2^\prime & , & Z_4 &), \\
     {\rm II} & (& {\bf 1} & , & Z_4 &), \\
     {\rm III} & (& Z_2^\prime & , & Z_4^\prime &), \\
     {\rm IV} & (& {\bf 1} & , & Z_4^\prime &). 
\ea
\ee
Calling the right handed $Z_4$ twist matrices $\th_4$ and $\th_4^\prime$,
we define the explicit $Z_2$ part of the $Z_2\times Z_4$ orbifold by 
$\th_2 = \th_4^3 \th_4^\prime$. The $Z_2$ action on left-movers is again
the outer automorphism of $E_8\times E_8$.

Models V through VIII are defined in the same way and again on $SO(12)$
lattices but with two modifications. The gauge shift vector is changed to
\be
\ba{c}
   V_4 = (+{1\over 4}, +{1\over 4}, -{1\over 2}, 0^5),
\ea
\ee
and since now $E_{vac}^{gauge} = 3/8$, one must also change the left
space twist to a reflection of either all six left coordinates ($Z_2^\ds$)
or only two ($Z_2^\tp$). Again there are four twist eigenvalue structures,
\be
\ba{lrcccl}
     {\rm V} & (& Z_2^\ds & , & Z_4 &), \\
     {\rm VI} & (&  Z_2^\tp & , & Z_4 &), \\
     {\rm VII} & (& Z_2^\ds & , & Z_4^\prime &), \\
     {\rm VIII} & (& Z_2^\tp & , & Z_4^\prime &).
\ea
\ee
The relevant NSR shift vectors $v$ corresponding to the $Z_4$, $Z_4^\prime$ 
and $Z_2$ twists are
\be
\ba{r}
   v_4        = (+{1\over 4}, +{1\over 4}, -{1\over 2}, \cero ), \vier
   v_4^\prime = (+{1\over 4}, -{1\over 4},    \cero   , \cero ), \vier
   v_4 - v_4^\prime = v_2 = ( \cero , +{1\over 2}, -{1\over 2}, \cero ).
\ea
\ee
As discussed before, one is advised to consider shifts $w$ in the $SO(12)$ 
lattice which are equivalent to the 
twists introduced above, namely
\be
\ba{r}
   w_4 =       (-{1\over 4}, +{1\over 4}, -{1\over 2}, +{1\over 2},0,0), \vier
   w_4^\prime= (-{1\over 4}, -{1\over 4},    \cero   , +{1\over 2},0,0), \vier
   w_4^\prime - w_4 = w_2 = (\cero ,-{1\over 2},+{1\over 2}, \cero ,0,0), \vier
   w_2^\prime = (-{1\over 2}, +{1\over 2}, \cero, \cero ,0,0), \vier
   w_2^\ds    = (+{1\over 2}, +{1\over 2}, +{1\over 2}, \cero ,0,0), \vier
   w_2^\tp    = (+{1\over 2}, \cero , \cero, \cero ,0,0). 
\ea
\ee
We may combine the left and right moving internal space parts, as well
as the NSR-fermions into 16 dimensional vector spaces with Lorentzian 
signature $(6,10)$. Then we can write space shift vectors
for the $Z_4$ and $Z_2$ suborbifolds, respectively, as
\be
\ba{llr}
\label{shifts}
   V_I     & = (w_2^\prime, w_4,        v_4)        &\;\; (- {1\over 2}), \vier
   V_{II}  & = (    0     , w_4,        v_4)        &\;\; (- 1),          \vier
   V_{III} & = (w_2^\prime, w_4^\prime, v_4^\prime) &\;\; (0),            \vier
   V_{IV}  & = (    0     , w_4^\prime, v_4^\prime) &\;\; (- {1\over 2}), \vier
   V_{V}   & = (w_2^\ds   , w_4,        v_4)        &\;\; (- {1\over 4}), \vier
   V_{VI}  & = (w_2^\tp   , w_4,        v_4)        &\;\; (- {3\over 4}), \vier
   V_{VII} & = (w_2^\ds   , w_4^\prime, v_4^\prime) &\;\; (+ {1\over 4}), \vier
   V_{VIII}& = (w_2^\tp   , w_4^\prime, v_4^\prime) &\;\; (- {1\over 4}), \vier
   V_2     & = (    0     , w_2,        v_2)        &\;\; (- 1),   
\ea
\ee
where the $Z_2$ is common to all eight models. I also displayed the length
squares of these vectors with respect to the Lorentzian signature. 
$V_2$ is chosen such that its scalar products with the other vectors vanish, 
thus avoiding extra complicated phases. 

The resulting spectra are displayed in Tables~\ref{spectra2} 
and~\ref{spectra3}. In the following, I will discuss the relevant
points to compute model VI.

\subsubsection*{Untwisted sector $(1,1)$}

The positive helicity ground states $h$ with their shift phases 
$e^{2 \pi i h v}$ are
\be
\ba{c|c|c|c}
     \th_R      &  Z_4     &   Z_2     \\ \hline
(+{1\over 2},+{1\over 2},+{1\over 2},+{1\over 2})&  1 &  1 \vier
(+{1\over 2},-{1\over 2},-{1\over 2},+{1\over 2})&  i &  1 \vier
(-{1\over 2},+{1\over 2},-{1\over 2},+{1\over 2})&  i & -1 \vier
(-{1\over 2},-{1\over 2},+{1\over 2},+{1\over 2})& -1 & -1
\ea
\ee 
Besides twist invariant adjoint representations of 
$E_6 \times SU(2) \times U(1)$ one finds states transforming as 
(i) $({\bf 27},{\bf 1})_{-2} + c.c.$, 
(ii) $({\bf 27},{\bf 2})_{+1} + ({\bf 1},{\bf 2})_{-3}$, and
(iii) $({\bf\ol{27}},{\bf 2})_{-1} + ({\bf 1},{\bf 2})_{+3}$, 
with $Z_4$ twist phases $-1$, $+i$ and $-i$, respectively.

\newpage

\begin{table}[h]
\begin {tabular}{|c|c|c|c|c|} \hline 
 
 & I & II & III & IV \\ \hline \hline $G_6$ &
$SO(8)\times SU(2)^2$ &
$SO(12)$ &
$SO(8)\times SU(2)^2$ &
$SO(12)$ \\ \hline & & & 

$({\bf 45},{\bf 1},{\bf 1},{\bf 1},{\bf 1})$ &
$({\bf 45},{\bf 1},{\bf 1})$ \\ & & &

$({\bf 1},{\bf 15},{\bf 1},{\bf 1},{\bf 1})$ &
$({\bf 1},{\bf 15},{\bf 1})$ \\ $(1,1)$ &

$({\bf 10},{\bf 6},{\bf 1},{\bf 1},{\bf 1})$ &
$({\bf 10},{\bf 6},{\bf 1})$ & & \\ &

$ 2 ({\bf 16},{\bf 4},{\bf 1},{\bf 1},{\bf 1})$ &
$ 2 ({\bf 16},{\bf 4},{\bf 1})$ &
$   ({\bf 16},{\bf 4},{\bf 1},{\bf 1},{\bf 1})$ &
$   ({\bf 16},{\bf 4},{\bf 1})$ \\ & & &

$(\ol{\bf 16},\ol{\bf 4},{\bf 1},{\bf 1},{\bf 1})$ &
$(\ol{\bf 16},\ol{\bf 4},{\bf 1})$ \\ \hline & & &

$({\bf 10},{\bf 1},{\bf 1},\un{{\bf 2},{\bf 1}})$ & \\ & 

$({\bf 1},{\bf 6},{\bf 1},\un{{\bf 2},{\bf 1}})$ & & 
$({\bf 1},{\bf 6},{\bf 1},\un{{\bf 2},{\bf 1}})$ & \\ $(1,\th_4)$ &

$({\bf 1},\ol{\bf 10},{\bf 1},\un{{\bf 2},{\bf 1}})$ & & & \\ & &

$ 2 ({\bf 16},\ol{\bf 4},{\bf 1})$ & &
$   ({\bf 16},\ol{\bf 4},{\bf 1})$ \\ & & & &

$(\ol{\bf 16},{\bf 4},{\bf 1})$ \\ \hline &

$({\bf 10},{\bf 1},{\bf 1},\un{{\bf 2},{\bf 1}})$ & & & \\ & 

$({\bf 1},{\bf 6},{\bf 1},\un{{\bf 2},{\bf 1}})$ & & 
$({\bf 1},{\bf 6},{\bf 1},\un{{\bf 2},{\bf 1}})$ & \\ $(\th_2,\th_4)$ & & &

$[({\bf 1}, {\bf 10},{\bf 1},\un{{\bf 2},{\bf 1}})]^+$ & \\ & & &

$[({\bf 1},\ol{\bf 10},{\bf 1},\un{{\bf 2},{\bf 1}})]^-$ & \\ & &

$[2 ({\bf 16},\ol{\bf 4},{\bf 1})]^+$ & &
$   ({\bf 16},\ol{\bf 4},{\bf 1})   $ \\ & &

$[2 (\ol{\bf 16},{\bf 4},{\bf 1})]^-$ & &
$   (\ol{\bf 16},{\bf 4},{\bf 1})   $ \\ \hline & &

$({\bf 54},{\bf 1},{\bf 1})$ & &
$({\bf 54},{\bf 1},{\bf 1})$ \\ & &

$({\bf 1},{\bf 20^\prime},{\bf 1})$ & &
$({\bf 1},{\bf 20^\prime},{\bf 1})$ \\ & &

$ 2 ({\bf 1},{\bf 1},{\bf 1})$ & &
$ 2 ({\bf 1},{\bf 1},{\bf 1})$ \\ $(1,\th_4^2)$ &

$({\bf 45},{\bf 1},{\bf 1},{\bf 1},{\bf 1})$ & & &
$({\bf 45},{\bf 1},{\bf 1})$ \\ &

$({\bf 1},{\bf 15},{\bf 1},{\bf 1},{\bf 1})$ & & &
$({\bf 1},{\bf 15},{\bf 1})$ \\ &

$({\bf 10},{\bf 6},{\bf 1},{\bf 1},{\bf 1})$ &
$({\bf 10},{\bf 6},{\bf 1})$ &
$ 2 ({\bf 10},{\bf 6},{\bf 1},{\bf 1},{\bf 1})$ & \\ \hline &

$({\bf 45},{\bf 1},{\bf 1},{\bf 1},{\bf 1})$ & & &
$({\bf 45},{\bf 1},{\bf 1})$ \\ &

$({\bf 1},{\bf 15},{\bf 1},{\bf 1},{\bf 1})$ & & &
$({\bf 1},{\bf 15},{\bf 1})$ \\ $(\th_2,\th_4^2)^+$ & &

$({\bf 10},{\bf 6},{\bf 1})$ & 
$({\bf 10},{\bf 6},{\bf 1},{\bf 1},{\bf 1})$ & \\ &

$(\ol{\bf 16},\ol{\bf 4},{\bf 1},{\bf 1},{\bf 1})$ &
$({\bf 16},{\bf 4},{\bf 1})$ &
$(\ol{\bf 16},\ol{\bf 4},{\bf 1},{\bf 1},{\bf 1})$ &
$({\bf 16},{\bf 4},{\bf 1})$ \\ &

$({\bf 1},{\bf 1},{\bf 1},{\bf 2},{\bf 2})$ &
$({\bf 1},{\bf 1},{\bf 12})$ &
$({\bf 1},{\bf 1},{\bf 1},{\bf 2},{\bf 2})$ &
$({\bf 1},{\bf 1},{\bf 12})$ \\ \hline & &

$({\bf 45},{\bf 1},{\bf 1})$ &
$({\bf 45},{\bf 1},{\bf 1},{\bf 1},{\bf 1})$ & \\ & &

$({\bf 1},{\bf 15},{\bf 1})$ &
$({\bf 1},{\bf 15},{\bf 1},{\bf 1},{\bf 1})$ & \\ $(\th_2,\th_4^2)^-$ &

$({\bf 10},{\bf 6},{\bf 1},{\bf 1},{\bf 1})$ & & &
$({\bf 10},{\bf 6},{\bf 1})$ \\ & 

$({\bf 16},{\bf 4},{\bf 1},{\bf 1},{\bf 1})$ &
$(\ol{\bf 16},\ol{\bf 4},{\bf 1})$ &
$({\bf 16},{\bf 4},{\bf 1},{\bf 1},{\bf 1})$ &
$(\ol{\bf 16},\ol{\bf 4},{\bf 1})$ \\ &

$({\bf 1},{\bf 1},{\bf  8},{\bf 1},{\bf 1})$ & &
$({\bf 1},{\bf 1},{\bf  8},{\bf 1},{\bf 1})$ & \\ \hline

& & & & \\

$(\th_2,1)^+$ &
same as $(\th_2,\th_4^2)^-$ &
same as $(\th_2,\th_4^2)^+$ &
$c.c.$ of $(\th_2,\th_4^2)^-$ &
$c.c.$ of $(\th_2,\th_4^2)^+$ \\ 

& & & & \\ \hline

& & & & \\

$(\th_2,1)^-$ &
same as $(\th_2,\th_4^2)^+$ &
same as $(\th_2,\th_4^2)^-$ &
$c.c.$ of $(\th_2,\th_4^2)^+$ &
$c.c.$ of $(\th_2,\th_4^2)^-$ \\ 

& & & & \\ \hline
\end {tabular}
\caption{Models from asymmetric $Z_2 \times Z_4$ orbifolds with 
         gauge group $[SO(10) \times SU(4)]^{k=2} \times G_6^{k=1}$.
         Superscripts $\pm$ refer to positive and
         negative discrete torsion.} 
\label{spectra2}
\end {table}

\newpage

The projection onto $Z_2$ invariant states is simple, because one can
always keep either the symmetric or the antisymmetric combination of
two $E_8$ vectors. Thus, we find the untwisted matter representations
$$2({\bf 27},{\bf 2})_{+1}+2({\bf 1},{\bf 2})_{-3}
   +({\bf 27},{\bf 1})_{-2}+ ({\bf\ol{27}},{\bf 1})_{+2}, $$
which transfrom trivially under the enhanced gauge group 
$G_6 = SO(10) \times U(1)$. As can be seen from the table, extra matter 
transforming under $G_6$ and invariant under $Z_4$ (related to the last 
helicity state) does not survive $Z_2$ projection.

\subsubsection*{First $Z_4$ twisted sector $(1,\th_4)$}
 
The only massless spinor in this sector is
\be
\ba{c}
           p = (-{1\over 4}, -{1\over 4}, \cero, +{1\over 2}).
\ea
\ee
The number of right and left fixed points is $N^f_R = 16$ and $N^f_L =  4$,
respectively, and $\det g_{inv} = 4$ so that we find a degeneracy $D = 4$.
In the shift description, however, the ground states are characterized 
by 8 right space vectors,
\be
\ba{c}
q_{1/2}=(-{1\over 4},+{1\over 4},\mp{1\over 2},\pm{1\over 2},\cero,\cero),\vier
q_{3/4}=(-{1\over 4},+{1\over 4},\pm{1\over 2},\pm{1\over 2},\cero,\cero),\vier
q_{5/6}=(+{1\over 4},-{1\over 4},\cero,\cero,\mp{1\over 2},\pm{1\over 2}),\vier
q_{7/8}=(+{1\over 4},-{1\over 4},\cero,\cero,\pm{1\over 2},\pm{1\over 2}),
\ea
\ee
and they are correlated with the left movers. Indeed, in order to pass
from $q_{1/2}$ to $q_{3/4}$ one must use an $SO(12)$ vector in both, the
left and right parts which corresponds to a root of $SO(24)$. 
Similarly, to pass from $q_{1/2}$ to $q_{5/6}$ one must use $SO(12)$
spinors on both sides. This corresponds to the spinor congruency class
in $Spin(24)/Z_2$ which is the Euclideanized Narain lattice of the
space part, i.e. the lattice with metric $G$ from Eq.~(\ref{bigg}).

As for the gauge part we have to look for states satisfying
\be
 {(P_{E_8\times E_8} + V^{gauge}_4)^2\over 2} = {7\over 8} \;\; {\rm or} \;\;
 {3\over 8}.
\ee
Hence, before $Z_2$ projection there are massless states
\be \label{x1} \ba{r} 
   [({\bf 27},{\bf 1})_{-1/2}({\bf 1},{\bf 1})_{+3/2} + 
     E_8 \leftrightarrow E_8 + 
            ({\bf 1},{\bf 2})_{-3/2}({\bf 1},{\bf 2})_{-3/2}] \otimes 
        [(r_{1},q_{1/2},p) + (r_{2},q_{3/4},p)], \vier
        ({\bf 1},{\bf 1})_{+3/2}({\bf 1},{\bf 1})_{+3/2} \otimes 
   [(r_3,q_{1/2},p) + (r_4,q_{3/4},p) +(r_5,q_{5/6},p) + (r_6,q_{7/8},p)], 
\ea \ee
where
\be
\ba{rcl}
   r_{1/2} &=& (\pm {1\over 2},\cero,\cero,\cero,\cero,\cero), \vier   
   r_3     &=& (  - {1\over 2},\ul{\pm 1,\cero,\cero,\cero,\cero}), \vier
   r_4     &=&(  + {1\over 2},\ul{\pm 1,\cero,\cero,\cero,\cero}), \vier
   r_{5/6} &=&(\cero,\pm{1\over 2},\pm {1\over 2},\pm{1\over 2},\pm{1\over 2},
                  \pm{1\over 2}) \;\;\; ({\rm even/odd \; \# \; of - signs}).
\ea
\ee
The first line (\ref{x1}) shows the fourfold twist vacuum degeneracy.
The second line comprises 104 states. In the twist formalism they would
arise through two half-oscillator exitations and 24 weights 
(${\bf 8}_v + {\bf 8}_s +{\bf 8}_c$) of an invariant $SO(8)$ times the 
fourfold vacuum degeneracy.

For the $Z_2$ projection, we first consider the phases 
$e^{-2\pi i (q_{1/2} w_2 + p v_2)} = - e^{-2\pi i (q_{3/4} w_2 + p v_2)} =
\mp 1$. The important point here is that both signs have to be used giving 
rise to both the symmetric {\em and\/} antisymmetric combinations of 
$E_8 \times E_8$, and regardless of an extra discrete torsion sign there are 
the states transforming under 
$E_6 \times SU(2) \times U(1) \times SO(10) \times U(1)$ like
$$[2({\bf 27},{\bf 1})_{+1}+({\bf 1},{\bf 3})_{-3}+({\bf 1},{\bf 1})_{-3}]
  {\bf 1}_{\pm 1/2} + ({\bf 1},{\bf 1})_{+3} {\bf 10}_{\pm 1/2}.$$
But there are also the states involving $q_{5/6}$ and $q_{7/8}$. For all of 
them we find $e^{-2\pi i (q w_2 + p v_2)} = +1$. That is,
if there is no further torsion sign all these states survive and give rise to
$$ 2 ({\bf 1},{\bf 1})_{+3} [{\bf 16}_0+{\bf\ol{16}}_0],$$
but in case of negative torsion these states are completely projected out.

\subsubsection*{$Z_2\times Z_4$ twisted sector $(\th_2,\th_4)$}

Massless spinors in this sector are
\be
\ba{c}
      p_\pm = (-{1\over 4}, +{1\over 4}, \mp {1\over 2}, \pm {1\over 2}),
\ea
\ee
and we have $N^f_R = N^f_L = \det g_{inv} = 4$, yielding $D = 2$.
The corresponding ground states are characterized by
\be
\ba{c}
      q_{1/2} = (-{1\over 4}, -{1\over 4}, \cero , \pm {1\over 2},0,0),
\ea
\ee
and gauge vectors (of the diagonal $E_8$) must satisfy
\be
   {(P_{E_8} + 2 V_4)^2\over 4} = {3\over 8}.
\ee
States satisfying the masslessness condition are (before $Z_2$ projection)
$$[({\bf 27},{\bf 1})_{+1}+({\bf 1},{\bf 1})_{-3} + c.c.] \otimes
[(r_1,q_1,p_\pm) + (r_2,q_2,p_\pm)].$$
The $Z_2$ projection requires great care. A look at the partition function
reveales the following phases: (i) an overall minus sign from the left-handed
vacuum energy, $E_{vac}^{gauge} = 1/2$, arising from the permutation of the 
two $E_8$ factors, so that
$e^{2\pi i E_{vac}^{gauge}} = -1$; (ii) another overall minus sign from the 
space shift, $e^{-i \pi V_2^2} = -1$; (iii) in case of negative discrete
torsion, yet another overall minus sign; (iv) ground state contributions
$e^{-2\pi i (q w_2 + p_\pm v_2)} = \mp 1$ for both $q$; and (v) another
possible sign from $e^{i\pi P_{E_8}^2/2}$ which contributes since
the dual of the invariant (diagonal) $E_8$ lattice is an odd lattice. 
The survivers are given by
$$ \ba{c} [({\bf 27},{\bf 1})_{+1}+({\bf 1},{\bf 1})_{-3}]\otimes 
[(r_1,q_1,p_+) + (r_2,q_2,p_+)], \vier
[({\bf\ol{27}},{\bf 1})_{-1}+({\bf 1},{\bf 1})_{+3}]\otimes 
[(r_1,q_1,p_-) + (r_2,q_2,p_-)]. \ea $$
States carrying the negative helicity vector $p_-$ must be complex conjugated. 
Thus we find
$$ 2[({\bf 27},{\bf 1})_{+1}+({\bf 1},{\bf 1})_{-3}], $$
while in case of negative torsion we would have the complex conjugate 
representations. 

\subsubsection*{Second $Z_4$ twisted sector $(1,\th_4^2)$}

This is the only twist sector without discrete torsion.
The massless spinors are
\be
\ba{c}
      p_\pm = ( \cero , \cero , \mp {1\over 2}, \pm {1\over 2}),
\ea
\ee
and we have $N^f_R = 16$, $N^f_L = 1$ and $\det g_{inv} = 4$, yielding $D = 2$
and corresponding to
\be
\ba{c}
      q_{1/2} = (\mp {1\over 2}, \pm {1\over 2}, \cero , \cero ,0,0), \vier
      q_{3/4} = (\pm {1\over 2}, \pm {1\over 2}, \cero , \cero ,0,0).
\ea
\ee
These we have to combine with states satisfying
\be
 {(P_{E_8\times E_8} + 2 V^{gauge}_4)^2\over 2} = 1\;\;{\rm or}\;\;{1\over 2},
\ee
and before further projections we have
$$ \ba{r}
({\bf 1},{\bf 2})_0[({\bf 27},{\bf 1})_{+1} + ({\bf 1},{\bf 1})_{-3} + c.c]
\otimes (0,q_{3/4},p_\pm), \vier
[({\bf 27},{\bf 1})_{+1} + ({\bf 1},{\bf 1})_{-3} + c.c]({\bf 1},{\bf 2})_0
\otimes (0,q_{3/4},p_\pm), \vier
({\bf 1},{\bf 2})_0 ({\bf 1},{\bf 2})_0 \otimes (r_{7/8},q_{1/2},p_\pm). \ea $$
In twist language,
\be
\ba{c}
   r_7 = (\pm 1,\cero,\cero,\cero,\cero,\cero), \vier
   r_8 = (\cero,\ul{\pm 1,\cero,\cero,\cero,\cero}),
\ea
\ee
would be described by vector weights (the {\bf 12}) of $SO(12)$.

Next we have to perform the $Z_4$ projection. It results a trivial overall 
twist phase due to $e^{-2\pi i (V_{VI}^2 + {V_4^{gauge}}^2)} = + 1$. We just 
need to consider the positive helicity vector since this is a twist sector of 
order two and the negative helicity vector gives simply the CPT partners. 
The relevant phases are $e^{-2\pi i (q_{1/2} w_4 + p_+ v_4)} = \mp 1$,
$e^{-2\pi i (q_{3/4} w_4 + p_+ v_4)} = - i$ and
$e^{2\pi i r_{7/8} w_2^\tp} = \mp 1$. The $Z_4$ survivors are given by
$$ \ba{c}
({\bf 1},{\bf 2})_0[({\bf 27},{\bf 1})_{+1} + ({\bf 1},{\bf 1})_{-3}]
\otimes (0,q_{3/4},p_+), \vier
[({\bf 27},{\bf 1})_{+1} + ({\bf 1},{\bf 1})_{-3}]({\bf 1},{\bf 2})_0
\otimes (0,q_{3/4},p_+), \vier
({\bf 1},{\bf 2})_0 ({\bf 1},{\bf 2})_0 
\otimes [(r_7,q_1,p_+) + (r_8,q_2,p_+)]. \ea $$

Finally, we turn to the $Z_2$ projection. Only the phase
$e^{-2\pi i (q_{1/2} w_2 + p_+ v_2)} = \pm 1$ is of interest here
which tells us to take the (anti)symmetric combination of $E_8$ vectors
for states involving $r_7$ ($r_8$). Hence, the contribution from this sector is
$$ \ba{c} 
2({\bf 27},{\bf 2})_{+1} {\bf 1}_0 + 2({\bf 1},{\bf 2})_{-3}{\bf 1}_0,
\vier ({\bf 1},{\bf 3})_0 {\bf 1}_{\pm 1} + ({\bf 1},{\bf 1})_0 {\bf 10}_0.
\ea $$

\subsubsection*{$Z_2\times Z_4^2$ twisted sector $(\th_2,\th_4^2)$}

Massless spinors in this sector are
\be
\ba{c}
      p_\pm = ( \cero , \mp {1\over 2}, \cero , \pm {1\over 2}),
\ea
\ee
and again we have $D=2$ corresponding to
\be
\ba{c}
      q_{1/2} = (\mp {1\over 2}, \cero , \pm {1\over 2}, \cero ,0,0),\vier
      q_{3/4} = (\pm {1\over 2}, \cero , \pm {1\over 2}, \cero ,0,0).
\ea
\ee
These vectors have to be combined with the solutions of 
\be
   {P_{E_8}^2\over 4} = {1\over 2} \;\; {\rm or} \;\; 0,
\ee
as well as with the half-integer oscillators.
These states comprise the full {\bf 248} of $E_8$ which has to be 
appropriately decomposed and combined with
$(0,q_{3/4},p_\pm)$, and there is also the {\bf 1} to be combined with 
$(r_{7/8},q_{1/2},p_\pm)$.

As for the $Z_4$ projection, there is as in the previous sector the trivial 
contribution from $e^{-2\pi i (V_{VI}^2 + {V_4^{gauge}}^2)} = + 1$,
but here is also a possible
torsion sign. Using $e^{-2\pi i (q_{3/4} w_4 + p_+ v_4)} = -1/-i$
and the various twist phases of the states inside the {\bf 248} as in the
untwisted sector, we find for positive torsion
$$ \ba{c} ({\bf 27},{\bf 1})_{-2} {\bf 1}_0 + c.c., \vier
({\bf 27},{\bf 2})_{+1} {\bf 1}_0 + ({\bf 1},{\bf 2})_{-3}{\bf 1}_0, \ea $$
while for negative torsion we would have
$$ \ba{c} ({\bf 78},{\bf 1})_0 {\bf 1}_0 + ({\bf 1},{\bf 3})_0 {\bf 1}_0 
+ ({\bf 1},{\bf 1})_0 {\bf 1}_0, \vier
({\bf\ol{27}},{\bf 2})_{-1} {\bf 1}_0 + ({\bf 1},{\bf 2})_{+3}{\bf 1}_0. \ea $$
Since $e^{-2\pi i (q_{1/2} w_4 + p_+ v_4)} = i/1$ and
$e^{2\pi i (r_{7/8} w_2^\tp)} = \mp 1$, we may use the states involving $q_2$
to infer for positive torsion extra states involving $r_8$,
$$({\bf 1},{\bf 1})_0 {\bf 10}_0,$$
while for negative torsion we would have instead the ones involving $r_7$,
$$({\bf 1},{\bf 1})_0 {\bf 1}_{\pm 1}.$$

\subsubsection*{$Z_2$ twisted sector $(\th_2,1)$}

This sector is very similar to the previous one, and the reader may
follow the same steps when knowing that the only possible overall $Z_4$ 
phase is the possible discrete torsion sign. For the model at hand,
it turns that the states arising from this sector
are identical to the ones from the previous sector.

\section{Discussion}
\label{summary}

The model computed in section~\ref{so10} with negative discrete torsion turns
out to have four generations and two adjoint representations. 
Phenomenologically, supersymmetric 
four generation models are not strictly ruled out, if one allows one
neutrino to be quite different and heavier compared to the others. Such
models have been constructed in References~\cite{Howie}. 

The obtained spectra for $E_6$ models from $Z_2\times Z_3$ orbifolds
are summarized in Table \ref{spectra}.
Surprisingly, models A and C, as well as models B and D turn out to be 
mirror models of each other. But the various states are rearranged
between the different sectors, and in particular, the adjoint representations
come from the untwisted sector in one model and from the third twisted 
(order two) sector
in the mirror. This is interesting because it shows that it is irrelevant
for phenomenology from which sector the adjoint Higgses arise. 
There is one model with two adjoint representations. It has vanishing 
net generation number without being non-chiral with respect to all
gauge groups. There is one model with 18 generations, 6 antigenerations and 
no exotic matter, which is an encouraging result as it shows that one can
have adjoint representations without extra exotics also for groups different
from $E_6$. 
Finally, there is one model with 9 generations, 3 antigenerations and sextets
of $SU(3)$. Although these models are related to the $Z_6$ ($Z_6^\prime$)
orbifold with 24 generations, we see that by using symmetric embedding and 
promoting it to level 2, the net generation number decreased by factors of 
two and four in the non-trivial cases. 

The obtained $SO(10)$ spectra are summarized in Table~\ref{spectra2}.
Inspection of the Table shows that models II and IV with negative torsion
are equivalent. Also model I with either torsion is equivalent to model
III$^+$, where the superscript denotes the torsion sign. The mirror model
of those is given by model III$^-$. Hence, there are 4 physically 
distinguishable models. These equivalences lead to an important 
observation: some of the adjoint representations arose as the twist
survivors inside a {\bf 248} of $E_8$. Others resulted as the 
antisymmetric combination in the product of two vectors of $SO(10)$.
The former are known to correspond to flat directions in the 
effective field theory. For example, if they are untwisted adjoint
fields, then they are easily seen to correspond to continuous Wilson lines.
But due to the equivalences just enumerated\footnote{Here I implicitly
assume that the equivalences persist at the massive levels and also for
the interactions. Although this seems reasonable for models constructed at 
maximally enhanced symmetry points, there is at least one example where two 
string models have the same massless spectra but differ at the massive levels:
the heterotic $SO(32)$ theory and Type I superstrings, where in the latter case
the spinor class of $SO(32)$ is pertubatively missing. Non-pertubatively, 
however, these theories are believed to be equivalent.}, 
the same conclusion must hold also for the latter type of adjoints. 

II$^+$ has 32 generations, 24 {\bf 10} and a {\bf 54} of SO(10), but
no adjoints. II$^-$ has 2 adjoints, 12 decouplets and a {\bf 54} but net 
generation number zero. IV$^+$ has even 4 adjoints and at the same time
a {\bf 54}, but also 
vanishing net generation number, as well as no {\bf 10}. 
The most interesting case is represented by models I and III with 8 net 
generations, 2 adjoints, 22 decouplets and no {\bf 54}. However, none
of the above spectra looks phenomenologically promising. Nevertheless,
it is noteworthy that in this class of models the appearance of multiple 
adjoint representations is rather generic. This is to be compared
with symmetric orbifolds where only one GUT Higgs field of $SO(10)$, either a 
{\bf 54} or {\bf 45} can be obtained~\cite{AFIU}. 

Cancellation of anomalies in the models of Tables~\ref{spectra} 
and~\ref{spectra2} can be checked with help of the relation\footnote{The 
symbol $\tr$ without specification refers to the trace in the
fundamental representation, while $\Tr$ means trace in the adjoint.} for 
two-index symmetric representations of $SU(N)$,
\be
  \tr_{s^{ij}} F_{SU(N)}^3 = (N+4) \tr F_{SU(N)}^3 \quad\quad\quad (N \geq 3),
\ee
which means for $SU(3)$ ($SU(4)$) that a {\bf 6} ({\bf 10}) representation 
contributes to the anomaly 7 (8) times the amount of a fundamental 
{\bf 3} ({\bf 4}). 

The obtained $E_6$ spectra from $Z_2\times Z_4$ orbifolds
are summarized in Table~\ref{spectra3}.

\begin{table}[t]
\begin {tabular}{|c|c|c|c|c|} \hline 
 
 & V & VI & VII & VIII \\ \hline \hline $G_6$ &
$SU(4)\times SU(4)$ &
$SO(10)\times U(1)$ &
$SU(4)\times SU(4)$ &
$SO(10)\times U(1)$ \\ \hline & & &

$({\bf 78},{\bf 1})_0$ &
$({\bf 78},{\bf 1})_0$ \\ $(1,1)$ &

$({\bf 27},{\bf 1})_{-2} + c.c.$ &
$({\bf 27},{\bf 1})_{-2} + c.c.$ &
$({\bf  1},{\bf 3})_0 + ({\bf  1},{\bf 1})_0$ &
$({\bf  1},{\bf 3})_0 + ({\bf  1},{\bf 1})_0$ \\ &

$ 2 ({\bf 27},{\bf 2})_{+1}$ &
$ 2 ({\bf 27},{\bf 2})_{+1}$ &
$   ({\bf 27},{\bf 2})_{+1} + c.c.$ &
$   ({\bf 27},{\bf 2})_{+1} + c.c.$ \\ &

$ 2 ({\bf  1},{\bf 2})_{-3}$ &
$ 2 ({\bf  1},{\bf 2})_{-3}$ &
$   ({\bf  1},{\bf 2})_{\pm 3}$ &
$   ({\bf  1},{\bf 2})_{\pm 3}$ \\ \hline & &

$ 2 ({\bf 27},{\bf 1})_{+1}  {\bf 1}_{\pm 1/2}$ & &
$   ({\bf 27},{\bf 1})_{+1}  {\bf 1}_{\pm 1/2} + c.c.$ \\ & 

$ 2 ({\bf  1},{\bf 2})_0 ({\bf 4} + {\bf\ol{4}},{\bf 1})$ & &
$   ({\bf  1},{\bf 2})_0 ({\bf 4} + {\bf\ol{4}},{\bf 1})$ & \\ $(1,\th_4)$ &

$ 2 ({\bf  1},{\bf 2})_0 ({\bf 1},{\bf 4} + {\bf\ol{4}})$ & &
$[2 ({\bf  1},{\bf 2})_0 ({\bf 1},{\bf 4} + {\bf\ol{4}})]^-$ & \\ & &

$   ({\bf  1},{\bf 3} + {\bf 1})_{-3}  {\bf 1}_{\pm 1/2}$ & & 
$   ({\bf  1},{\bf 1})_{-3}  {\bf 1}_{\pm 1/2} + c.c.$ \\ & &

$   ({\bf  1},{\bf 1})_{+3}  {\bf 10}_{\pm 1/2}$ & & \\ & &
 
$[2 ({\bf  1},{\bf 1})_{+3} {\bf 16}_0]^+$ & & \\ & &

$[2 ({\bf  1},{\bf 1})_{+3} {\bf\ol{16}}_0]^+$ & & \\ \hline & &

$ 2 ({\bf 27},{\bf 1})_{+1}  {\bf 1}_{\pm 1/2}$ & &
$   ({\bf 27},{\bf 1})_{+1}  {\bf 1}_{\pm 1/2} + c.c.$ \\ &

$   ({\bf  1},{\bf 2})_0 ({\bf 4} + {\bf\ol{4}},{\bf 1})$ & &
$ 2 ({\bf  1},{\bf 2})_0 ({\bf 4} + {\bf\ol{4}},{\bf 1})$ & \\ 
                                                 $(\th_2,\th_4)^+$ & & & &

$   ({\bf  1},{\bf 3})_{+3}  {\bf 1}_{\pm 1/2}$ \\ & &

$ 2 ({\bf  1},{\bf 1})_{-3}  {\bf 1}_{\pm 1/2}$ & & 
$   ({\bf  1},{\bf 1})_{-3}  {\bf 1}_{\pm 1/2}$ \\ & & & &

$   ({\bf  1},{\bf 1})_{-3}  {\bf 10}_{\pm 1/2}$ \\ \hline

& & & & \\

$(\th_2,\th_4)^-$ & 
same as $(\th_2,\th_4)^+$ &
$c.c.$ of $(\th_2,\th_4)^+$ &
same as $(\th_2,\th_4)^+$ &
$c.c.$ of $(\th_2,\th_4)^+$ \\

& & & & \\ \hline &

$ 2 ({\bf\ol{27}},{\bf 2})_{-1}$ &
$ 2 ({\bf 27},{\bf 2})_{+1}$ &
$   ({\bf 27},{\bf 2})_{+1} + c.c.$ &
$   ({\bf 27},{\bf 2})_{+1} + c.c.$ \\ $(1,\th_4^2)$ &
 
$ 2 ({\bf  1},{\bf 2})_{+3}$ &
$ 2 ({\bf  1},{\bf 2})_{-3}$ &
$   ({\bf  1},{\bf 2})_{\pm 3}$ &
$   ({\bf  1},{\bf 2})_{\pm 3}$ \\ &

$({\bf  1},{\bf 3})_0 ({\bf 1},{\bf 6})$ &
$({\bf  1},{\bf 3})_0  {\bf 1}_{\pm 1}$ &
$({\bf  1},{\bf 3})_0 ({\bf 1},{\bf 6})$ &
$({\bf  1},{\bf 3})_0  {\bf 1}_{\pm 1}$ \\ &

$({\bf  1},{\bf 1})_0 ({\bf 6},{\bf 1})$ &
$({\bf  1},{\bf 1})_0  {\bf 10}_0$ &
$({\bf  1},{\bf 1})_0 ({\bf 1},{\bf 6})$ &
$({\bf  1},{\bf 1})_0  {\bf 1}_{\pm 1}$ \\ \hline &

$({\bf 78},{\bf 1})_0$ & &
$({\bf 78},{\bf 1})_0$  & \\ &

$({\bf  1},{\bf 3})_0 + ({\bf 1},{\bf 1})_0$ &
$({\bf 27},{\bf 1})_{-2} + c.c.$ &
$({\bf  1},{\bf 3})_0 + ({\bf 1},{\bf 1})_0$ &
$({\bf 27},{\bf 1})_{-2} + c.c.$ \\ $(\th_2,\th_4^2)^+$ &

$   ({\bf\ol{27}},{\bf 2})_{-1}$ &
$   ({\bf 27},{\bf 2})_{+1}$ &
$   ({\bf 27},{\bf 2})_{+1}$ &
$   ({\bf\ol{27}},{\bf 2})_{-1}$ \\ &

$   ({\bf  1},{\bf 2})_{+3}$ &
$   ({\bf  1},{\bf 2})_{-3}$ &
$   ({\bf  1},{\bf 2})_{-3}$ &
$   ({\bf  1},{\bf 2})_{+3}$ \\ &

$({\bf  1},{\bf 1})_0 ({\bf 6},{\bf 1})$ &
$({\bf  1},{\bf 1})_0  {\bf 10}_0$ &
$({\bf  1},{\bf 1})_0 ({\bf 1},{\bf 6})$ &
$({\bf  1},{\bf 1})_0  {\bf 1}_{\pm 1}$ \\ \hline & &

$({\bf 78},{\bf 1})_0$ & &
$({\bf 78},{\bf 1})_0$ \\ &

$({\bf 27},{\bf 1})_{-2} + c.c.$ &
$({\bf  1},{\bf 3})_0 + ({\bf 1},{\bf 1})_0$ &
$({\bf 27},{\bf 1})_{-2} + c.c.$ &
$({\bf  1},{\bf 3})_0 + ({\bf 1},{\bf 1})_0$ \\ $(\th_2,\th_4^2)^-$ &

$   ({\bf 27},{\bf 2})_{+1}$ &
$   ({\bf\ol{27}},{\bf 2})_{-1}$ &
$   ({\bf\ol{27}},{\bf 2})_{-1}$ &
$   ({\bf 27},{\bf 2})_{+1}$ \\ &

$   ({\bf  1},{\bf 2})_{-3}$ &
$   ({\bf  1},{\bf 2})_{+3}$ &
$   ({\bf  1},{\bf 2})_{+3}$ &
$   ({\bf  1},{\bf 2})_{-3}$ \\ &

$({\bf  1},{\bf 1})_0 ({\bf 1},{\bf 6})$ &
$({\bf  1},{\bf 1})_0  {\bf 1}_{\pm 1}$ &
$({\bf  1},{\bf 1})_0 ({\bf 6},{\bf 1})$ &
$({\bf  1},{\bf 1})_0  {\bf 10}_0$ \\ \hline

& & & & \\

$(\th_2,1)^+$ &
same as $(\th_2,\th_4^2)^-$ &
same as $(\th_2,\th_4^2)^+$ &
$c.c.$ of $(\th_2,\th_4^2)^+$ &
$c.c.$ of $(\th_2,\th_4^2)^-$ \\ 

& & & & \\ \hline

& & & & \\ 

$(\th_2,1)^-$ &
same as $(\th_2,\th_4^2)^+$ &
same as $(\th_2,\th_4^2)^-$ &
$c.c.$ of $(\th_2,\th_4^2)^-$ &
$c.c.$ of $(\th_2,\th_4^2)^+$ \\ 

& & & & \\ \hline
\end {tabular}
\caption{Models from asymmetric $Z_2 \times Z_4$ orbifolds with 
         gauge group $[E_6\times SU(2) \times U(1)]^{k=2} \times G_6^{k=1}$.
         Superscripts are as in Table 3. Only non-trivial representations of
         $G_6$ are shown.}
\label{spectra3}
\end {table}

Similar to the $SO(10)$ models, model V with either torsion is equivalent 
to model VII$^-$, but VII$^+$ is different. These models have zero net 
generation number, namely $8+8$ and $6+6$ generations, and are non-chiral. 
Models VI and VIII with negative torsion are equivalent, while model VIII$^+$ 
represents the mirror. These models are the most interesting ones as they 
have four net $E_6$ generations ($13 + 9$) and two adjoints. Again model 
VI$^+$ is different and has $23 + 3$ generations, but no adjoint $E_6$ matter.

Models VI and VIII have an anomalous $U(1)$. In general, the anomaly is given 
by
\be
\label{anomaly2}
\ba{lll}
  (2 \pi)^2 I &=& {1\over 48} \tr R^2 \sum\limits_{i,A} s^i_A (q^i_A F_A)
       - {1\over 6} \sum\limits_{i,A} s_A^i (\tr_{R^i} F^3_A)
           - {1\over 2} \sum\limits_{i,j,A,B} s_{AB}^{ij}
             (\tr_{R^i} F^2_A) (q_B^j F_B) \acht
           &-& \sum\limits_{i,j,k,A,B,C} s_{ABC}^{ijk} 
            (q_A^i F_A) (q_B^j F_B) (q_C^k F_C),
\ea \ee
where $s^i_A$ is the number of multiplets transforming in representation 
$R^i$ (or with charge $q_A^i$) under group factor $G_A$, $s_{AB}^{ij}$ is the 
number of multiplets transforming in representation $(R^i,R^j)$ under 
$G_A \times G_B$, etc. The trace over curvature matrices 
in $R$ is in the vector representation of $SO(3,1)$. The second term is
the usual cubic anomaly, which must of course vanish for non-Abelian
group factors; for Abelian factors $\tr_{R^i}$ has to be replaced by 
$q_A^{i3}$ (and in the third term by $q_A^{i2}$).
Cancellation of anomalies then requires factorization into
\be 
  (2 \pi)^2 I = [\tr R^2 - \sum\limits_A k_A \alpha_A^{(1)} \Tr F_A^2]
     \times     [\sum\limits_B \alpha_B^{(2)} F_B],
\ee
with
\be \label{alf} \alpha^{(1)}_A = {1\over \tilde{h}_A}. \ee
For $U(1)$ factors (omitting the trace symbol)
\be \alpha^{(1)}_{U(1)} = N, \ee
where $N$ is defined through the level 1 relation,
\be
  h_{U(1)} = {q^2\over N},
\ee
and with the normalization suggested in Reference~\cite{6D} one would choose
$N=1$. While at level one $q^2$ is indeed directly related to the conformal
dimension, at higher levels one can still use this relation when (like in
the present case) the higher level $U(1)$ factor can be traced back to 
a level one $U(1)$. Putting everything together one may write
\be
  \alpha^{(2)}_{U(1)} = {1\over 48} \sum\limits_i q_i
\ee
and there is the condition
\be
\label{rat}
   {\sum\limits_i q_i^3\over \sum\limits_i q_i} = {k N\over 8}.
\ee
In the cases under consideration, $N=12$ and $k=2$ and Eq.~(\ref{rat}) can
be seen to be satisfied; moreover, $\alpha^{(2)}_{U(1)}=15$ (3) for model
VI with positive (negative) torsion. The mixed gauge anomalies can now
be checked using 
\be
   \sum\limits_{i,j} s^{ij}_{A,U(1)} q^i = 
                 2 k_A \alpha^{(1)}_A \alpha^{(2)}_{U(1)},
\ee
where $\alpha^{(1)}_A$ is given by Eq.~(\ref{alf}) when working with traces
in adjoint representations; when using fundamental representations, the
$\alpha^{(1)}_A$ are given by~\cite{6D},
\be \label{alp1} \ba{lclccl}
   \alpha^{(1)}_{SU(N)}  &=& \alpha^{(1)}_{Sp(N)}  &=& 2, & \vier
   \alpha^{(1)}_{SO(N)}  &=& \alpha^{(1)}_{G_2}    &=& 1, & (N \geq 5) \vier
   \alpha^{(1)}_{F_4}    &=& \alpha^{(1)}_{E_6}    &=& {1\over 3}, & \vier
&& \alpha^{(1)}_{E_7}    &=& {1\over 6}, & \vier
&& \alpha^{(1)}_{E_8}    &=& {1\over 30}.&
\ea \ee
Sometimes it is necessary to use Eq.~(\ref{casimir}). For example, for
the present cases one needs 
\be 
   \tr_{16} F_{SO(10)}^2 = 2\; \tr F_{SO(10)}^2.
\ee
Explicit relations are provided in the appendix of~\cite{6D}.

The net generation numbers of all 11 inequivalent models is even.
The relative difficulty to obtain odd generation numbers has been noted
before in the context of the free fermionic
construction~\cite{CCHL,Cleaver,Finnell}. In level 1 orbifolds, it is
known that turning on quantized Wilson lines can result in odd and, in
particular, three generations~\cite{IKNQ}. The construction introduced
in this paper possesses the option of turning on Wilson lines, as well,
and this represents one of the possible generalizations. Another important
generalization are models with levels $k$ larger than 2, obtained by permuting 
$k$ identical group factors. This way, one may obtain $[SO(10)]^{k=3}$ models
with a massless {\bf 120} multiplet. On the other hand, the fermionic 
construction allows only for levels of the form $k = 2^n$ with $n$ an integer.
As mentioned in the introduction, one may also attempt to construct
models with Standard Model gauge group at level 2 in order to improve
coupling unification~\cite{BFY}. In the construction at hand, this 
requires to go to higher twist orders. 

As a final spin off, the techniques developed
in this work can be used even for known models at level $k=1$: 
utilizing exclusively shifts as in Eqs.~(\ref{shifts}) for both, the space
and gauge parts\footnote{These are shifts acting in odd self-dual Lorentzian
lattices with signature $(22,10)$. Indeed, these models are explicit 
realizations of the ``covariant lattices'' as introduced in~\cite{LLS}.},
it is straightforward to compute correlation functions
for the popular three generation models with quantized Wilson lines mentioned 
above~\cite{IKNQ}. Basically, one would only have to evaluate
exponentials of the conformal field theory, similar to the torus case. 
In contrast, standard techniques~\cite{DFMS} would require the 
calculation of twist field correlation functions which is rather involved. 
Moreover, in the presence of quantized Wilson lines, which as discussed
in section~\ref{asymmetric} are related to asymmetric orbifolds, one would
need the technology outlined in Reference~\cite{NSV2}. This has not been
carried out successfully, so that for the most interesting class of orbifolds
interactions are presently unavailable. 

To conclude, I have introduced a new approach to construct higher level 
string models. The construction is based on orbifolds which is has the
advantage that the models are exactly soluble and 
allow for exact deformations using orbifold moduli. For example,
the untwisted adjoint Higgs fields can be represented as continuous
Wilson line moduli. Moreover, using asymmetric twists it is possible to avoid 
the ``one GUT Higgs theorem'' valid in symmetric orbifolds
with $SO(10)$ gauge groups~\cite{AFIU}.

\section*{Acknowledgements}

This research was supported in part 
by the Spanish Ministerio de Educaci\'on y Ciencia,
by the Department of Energy, and
by the National Science Foundation under Grant No.\ PHY94-07194.

\end{document}